%Paper: hep-th/9306090
%From: Samir Mathur <me@ctpdown.mit.edu>
%Date: Fri, 18 Jun 93 11:22:16 EDT

%%%%%%%%%%%%%%%%%%%%%%%%%%%%%%%%%
\input harvmac
\overfullrule=0 pt
\def\ad{a^{\dagger}}
\def\boxx{\bar \sqcup}

\Title{}
{{Is the Polyakov path integral prescription}}
\vskip -1.3 in
\Title{}{ too restrictive?}
\centerline{{\bf Samir D. Mathur}}

\bigskip\centerline{Center for Theoretical Physics}
\centerline{Massachussets Institute of Technology}
\centerline{Cambridge, MA 02139}
\vskip .7cm

In the first quantised description of
strings, we integrate  over target space co-ordinates
$X^\mu$ and world sheet metrics $g_{\alpha\beta}$.
Such  path integrals give scattering amplitudes
between the  `in' and `out' vacuua  for  a time-dependent
 target space geometry.  For a complete description of
 `particle creation' and the corresponding backreaction,
 we need instead the causal amplitudes obtained from an `initial value
formulation'.  We argue, using the analogy of
a scalar particle in
curved space, that  in the first quantised path integral one should integrate
over $X^\mu$ and
world sheet {\it zweibiens}. This extended formalism can be made to
yield causal amplitudes; it also naturally allows incorporation of density
matrices in a covariant manner.
(This paper is an expanded version of hep-th 9301044)

\vskip .1 in
\Date{\hfill  May 1993}

\newsec{Introduction.}

When we quantise a field in curved space, many interesting
phenomena arise. For example in an expanding Universe we have particle
creation, which means that a spacetime which looks empty in terms of
particle modes natural in the past, may look full of particles on using
co-ordinates natural to the future. The Minkowski vacuum appears to have a
particle flux for an accelerated observer. For black holes, imposing
vacuum conditions at past null infinity gives Hawking radiation in the
future, due to the time-dependent gravitational field of a collapsing
object.  (\ref\BIRREL{N.D. Birrell and P.C.W. Davies `Quantum fields in
curved space' (1982) Cambridge Univ. Press.} and references therein.)

For a complete description of the physics, we need also
the backreaction of the quantum field on the geometry.  At
this point we need a consistent theory of quantum gravity,
to deal with the gravity loops which arise. For this reason backreaction
questions have been traditionally restricted to semiclassical
order (apart from recent studies on low dimensional gravity
theories \ref\CGHS{C.G. Callan, S.B. Giddings, J.A. Harvey and
A. Strominger, Phys. Rev. {\bf  D 45} (1992) R1005.}).

Strings provide  a consistent theory
of quantised matter and gravity, so we would like
to investigate the issue of particle creation and backreaction in string
theory.  Several examples of strings in time-dependent
backgrounds were studied by de Vega  et. al \ref\DEVEGA{
H.J. de Vega,  Preprint PAR LPTHE 92/50  (hep-th 9302052), and
references therein.}.  In their first quantised treatment
of the string, they find that an existing
string may get excited to higher mode levels in moving through
the time-dependent geometry (e.g.  spacetime with a gravitational wave),
but it remains one string.  Thus at their tree
level calculation we do not see the
creation of a `flux of strings', which would
be the analogue of  the `creation of particles'  for
field theory  in curved spacetime.

In what way do we expect to see `created particles' in a first quantised
    language? Suppose a Universe starts in the vacuum, and undergoes
    expansion which creates particles.
Consider the two point propagator computed for  spacetime points
in the region where the `bath' of created particles is nonvanishing.We
may compare this situation to the computation of the propagator
at finite temperature, where also a bath of ambient particles exists.
In the latter case we know that the propagator depends on the temperature, and
with this modified propagator Feynman diagrams can be computed to develop
perturbation theory. With our `created particles'
there should be a similar modification of the propagator, except
that the `bath' would be neither exactly thermal nor time-independent
in general.

In this paper we take the simple case of a scalar field,  and show how a
careful handling of the `sum over paths' in the first quantised
propagator  allows us to obtain the modifications needed for
incorporating the effects of created particles.
 If we compute the analogue of the
Polyakov path integral for a particle, we get a two point
scattering amplitude between the `in' and `out' vacuum states.  Such a naively
computed first quantised
propagator does not directly `see' the
created particles. What we need instead is a propagator that reproduces the
causal amplitude obtained by taking the expectation value of field operators in
 a specified state.  Our `first quantised' calculation reproduces
this propagator. A density
matrix formalism develops naturally, and the particle fluxes produced
in time dependent geometries become  an essential part
of the first quantised description. It appears that only such an
extension of the Polyakov path integral can adequtely describe
the physics of quantised gravity plus matter.

For the string case we have the further issue of obtaining the background
field equations by a $\beta$-function calcaulation. The tree level
$\beta$-function gives  the classical vacuum; but we need the
higher string loop contributions to take into account the backreaction
of created particles. We argure that our modified propagator, rather
than the Polyakov prescription, should be used in  computing the loop
amplitudes.

 \bigskip
{\bf The issue.}

Consider the free massive scalar field.
For the `propagator' from one spacetime point $x_1$ to
another spacetime point $x_2$ , the analogue of the Polyakov
path integral gives \ref\GOVAERTS{J. Govaerts,
`Hamiltonian quantisation and constrained dynamics' (1991) Leuven Univ.
Press.}
 \eqn\ONEONEP{D(x_2,x_1)~=~\int_0^\infty d\lambda D[X]e^{-i\int_0^1
d\tau(1/2)(\dot X^2/\lambda+m^2\lambda)-\epsilon\lambda}}
 where $\tau$ parametrises the world line, and $X(0)=x_1$,
 $X(1)=x_2$. The action is multiplied by $i$ because we are
 in spacetime with Minkowski signature. The regulator term
 $-\epsilon \lambda$ ensures convergence of the path integral.

 What precisely {\it is} the answer we obtain? In spacetimes with Euclidean
 signature there is a unique Green's function, determined by the boundary
 condition that the amplitude to propagate to infinity falls to zero
 in all directions.  Such is not the case for spacetimes with Minkowski
 signature, where various Green's functions can be defined, differing
{} from each other by solutions of the homogeneous field equation.
 For describing the choice of boundary conditions implicit in
 \ONEONEP\ let us assume that the spacetime does not expand
 `too rapidly' in the far past ($t\rightarrow -\infty$) or in the far future
 ($t\rightarrow\infty$). Then we obtain naturally defined vacuum states
 for the Klein-Gordon field, $|0>_{in}$ and  $|0>_{out}$ for
 $t\rightarrow -\infty$ and $t\rightarrow \infty$ respectively. For a
 time-dependent geometry, in general  $|0>_{out}\ne |0>_{in}$. In this
 situation it is known \ref\RUMPF{H. Rumpf and H.K.
Urbantke, Ann. Phys. {\bf 114} (1978) 332, H. Rumpf, Phys. Rev. {\bf D 24}
(1981) 275, {\bf D 28} (1983) 2946.}
  that the RHS of \ONEONEP\  computes
\eqn\ONETWOP{D(x_2,x_1)~=~{
{}_{out}<0|T[\phi(z_2)\phi(z_1)]|0>_{in}\over {}_{out}<0|0>_{in}}~
\equiv~G_F(x_2,x_1)}
which is the analogue of the Feynman propagator in the time-dependent
geometry.

Such  `in-out' amplitudes are
relevant for computing scattering amplitudes where a finite number of `in'
particles scatter to a finite number of `out' particles. By
contrast, for an `initial value problem', with
the initial state the `in' vacuum, we would compute  amplitudes like
${}_{in}<0|T[\phi(z_2)\phi(z_1)]|0>_{in}$. In particular, we know
that for a source term to Einstein's equations we should use
`true expectation values' $<\psi|T_{\mu\nu}|\psi>$ for
the stress tensor, not `in-out'  vacuum amplitudes  \BIRREL . We
illustrate (in section 5 below) the difference between these two different
kinds of amplitudes by a simple example in $1+1$ spacetime.
We compute the spatial average of ${}_{in}<0|T_{\mu\nu}(\eta,x)|0>_{in}$
(called $<T_{\mu\nu}(\eta)>_{in~in}$) and of
${}_{out}<0|T_{\mu\nu}(\eta,x)|0>_{in}
/{}_{out}<0|0>_{in}$ (called $<T_{\mu\nu}(\eta)>_{in~out}$) in a Universe which
is in the initial vacuum and  undergoes sudden
expansion at $\eta=0$. ($\eta$ is the conformal time.)
For $\eta<0$, $<T_{00}>_{in~in}$ is just the
vacuum energy. For $\eta>0$ it becomes the vacuum energy
plus the energy of created particles. But  $<T_{00}>_{in~out}$
is just the vacuum energy  in both ranges of $\eta$. Worse,
 $<T_{11}>_{in~out}$ is {\it complex}, so it cannot be a
 source term for the gravitational field.

To compare `in-in' and `in-out' amplitudes  we note that ${}_{in}<0| =
C_0~{}_{out}<0|~+~C_2~{}_{out}<2|~+~C_4~{}_{out}<4|~+\dots$ where ${}_{out}<2|$
is a state
with one pair of `out' particles etc.
 The first quantised path integral  described above
 computed only
 `in-out' amplitudes.  This may not appear to be a serious
 shortcoming of the
 first quantised approach in the particle case
  because one could compute successively
 $ {}_{out}<0|0>_{in}$, ${}_{out}<2|0>_{in}$, etc. and thus reconstruct
 the result of the initial value problem. But this is not an adequte
 treatment for the string case, which was the motivation for
studying the first quantised approach in the first place.  With
strings we do not start with a Lagrangian governing the field;
instead we discover backgrounds satisfying the field equations
by demanding that string propagation in the background be
`consistent'. But which propagator should we demand consistency for:
the `in-in' one or the `in-out' one? These two choices would in general
give different answers for the background field (considering $\beta$-
functions to  one loop or beyond). The discussion above indicates
that   the  `in-in' propagator  is the  one which
correctly gives
 the stress
tensor of the created flux of `out' particles,  and is therefore
the one that should be involved in the multiloop contributions
to the $\beta$-function.

Can we develop a natural extension of the Polyakov prescription
so that we compute `in-in' amplitudes? In the second quantised
field language, `in-in' perturbation theory is described in
the `real time formalism' of many-body theory. We
outline this formalism below, and then give our results on how
this  formalism arises naturally and covariantly in the first-quantised
`sum over paths' language.

{\it Note}: In this paper we use the term `Polyakov path integral'
for any prescription where one sums over spacetime co-ordinates
on the world sheet /world line  and the metrics on
this world sheet/world line. This approach was developed primarily
for Euclidean target space metrics. We are assuming here
that the prescription of summing over co-ordinates and metrics has been
extended in some way to spacetime with Minkowski signature.
Thus the world sheet will also have Minkowski signature, and
construction of higher genus surfaces may require explicit use of interaction
vertices.

\bigskip
{\bf The real time formalism.}

To obtain a perturbation theory for the initial value problem,
one uses the `real time formalism', developed in the context of
time-dependent many body theory \ref\KEL{L.V. Keldysh, J.E.T.P. {\bf 20} (1965)
1018.}\ref\SCH{J.S. Schwinger, J.  Math. Phys. {\bf 2} (1961) 407.}. Suppose we
start in the
vacuum $|0>_{in}$ at $t=-\infty$. We wish to compute
 the expectation value of a product
of observables $O_i$, defined at times $t_i$, in the state  $|0>_{in}$.
 This expectation value
 can be expressed as  $\Tr\{\rho_0 O_1\dots O_n\}/\Tr\rho_0$, with
$\rho_0=|0>_{in}{}_{in}<0| $ the density matrix for a pure state. Because both
the bra and the ket states composing $\rho_0$ are given at $t=-\infty$, the
interaction Hamiltonian acts along the `time path' leading from
$t=-\infty$ to $t=\infty$, and then back to $t=-\infty$.  Correspondingly,
we need propagators from any time on the first or
second time path segment to any  other time on   either
path segment.  We collect these four propagators (appropriate
to the four possible choices of time path segments) into
a 2x2 {\it matrix} propagator. Vertices arising from the interaction
Hamiltonian acting on the first time path segment have
the usual  coupling $-ig$, while those on the second path segment
have coupling $ig$. With this extension of rules, the computation
of Feynman diagrams gives correlation functions for the
density matrix $\rho_0$, inserted at $t=-\infty$.

Note that the above formalism computed `in-in' correlation
functions, regardless of whether  or not the `in' vacuum differed
{}from the `out' vacuum. The density matrix $\rho_0$ can be
replaced by an arbitrary density matrix $\rho$. A special
class of $\rho$ emerge in the development of perturbation
theory:  density matrices expressible as `exponential of
quadratic in the field operator'. For such $\rho$ the Wick
decomposition holds: the expectation of a product of several operators
decomposes into sums of products of expectations for pairs
of operators. The propagator  of the real time formalism
encodes a choice of such a `special' density matrix. Deviations
of $\rho$ from this special class give `correlation kernels',
which are handled perturbatively
 just like interaction vertices.

The relevance of the real time formalism for field theory
in time dependent geometry is easy to see. Firstly,  cosmology
appears set up as an initial value problem, rather than a scattering
problem.  Secondly, `particle creation'  is  generally obtained as a
Bogoliubov
tranformation on a vacuum state. The particle flux produced this way
is described by a density matrix of the form `exponential of quadratic in
field'.  This flux is therefore precisely of the form that can be
incorporated into the propagator of the real time formalism, so that
backreaction computations can be examined by loop calculations
using the matrix propagator.  Thirdly, the density matrix of
 the real time formalism can be taken to describe a mixed state,
 whereupon one can study correlations in an evolving distribution
 of particles.  Note that the `periodic imaginary time' formalism
 for finite temperature correlations applies only to time independent
 thermal distribution functions. But in a theory with gravity,  a nonempty
 distribution of particles gives rise to a nonvanising stress tensor,
 which leads to a time dependence of the spacetime geometry in general,
 and a corresponding evolution of the  particle distribution.  Thus the
 real time formalism must be used in developing kinetic theory
 for a theory with gravity; the imaginary time formlism can only
 provide quasi-static approximations.

 The above description of the real time method pertained to
 the second quantised field language.  In order to apply the
 method to strings, we need to obtain it in a first quantised
 path integral language.  At first one might think that this
 would not be possible, since a first quantised path integral seems
 to describe the propagation of just one particle, while the
 real time formalism is built to handle fluxes.  Nevertheless,
 it turns out that one does find the entire structure of the formalism,
 when one carefully quantises the action of, say, a scalar particle.
 The phrase `first quantised' is not really correct for the
 resulting path integral; what one gets is a `proper time representation'
 for the real time propagator.\foot{The author thanks L. Ford for pointing
 out this issue of terminology.}
  The fact that one reaches this
 object after starting with what looks like just one particle,
 reflects the ambiguity of the particle concept in curved space.
 The Polyakov path integral (and its particle analogue)
 chooses a prescription to define the propagation amplitude
 such that this ambiguity is supressed.  In restoring the ambiguity,
 one regains the freedom to have an `in-in' formulation, and also
 the choice of an initial density matrix.

\bigskip
{\bf Approach and results.}

We start with the geometric action $m\int ds$ for a scalar
particle moving in a spacetime with Minkowski signature.
$ds=(dX^\mu dX_\mu)^{1/2}$ has a sign ambiguity. Along
a trajectory  contributing to the sum over paths, the path can
change several times between being timelike and being spacelike.
At each change $dX^\mu dX_\mu$ passes through zero,
and we have to make the choice of root afresh.  We might try to
evade the problem by passing to the quadratic form of the
action. One often thinks of this form of the path integral
as being a sum over functions $X^\mu(\tau)$ and over
metrics $g_{\tau\tau}(\tau)$ on the world line of the particle.
In fact this would be the analogue  for the
scalar particle of the Polyakov prescription. But in constraining the square
root action to
bring it to the quadratic form, what we really get is a sum over
the  $X^\mu(\tau)$  and over the {\it einbein} $\lambda(\tau)$
on the world line. Since the  einbein is the square root of the
metric, two signs of the einbein correspond to the same
$g_{\tau\tau}(\tau)$; this appears to be a reflection of the
sign ambiguity in the original square root action.

To evaluate the path integral we gauge fix $\lambda(\tau)$ to
a constant by using the freedom of diffeomorphisms along
the world line.  If the world line geometry were described by
the metric $g_{\tau\tau}(\tau)$, it is evident that the only degree of
freedom remaining of this geometry would be the length $\Lambda$
of the entire world line, and $\Lambda$ should be integrated
{}from $0$ to $\infty$.  But since we have instead a sum over
the einbein  $\lambda(\tau)$, the situation is more complicated.
Diffeomorphisms can fix  $\lambda$ to a positive constant in any
interval where $\lambda$ is positive, and to a negative
constant in any interval where it is negative.  These two
kinds of intervals will alternate along the world line for a typical
realisation  of  $\lambda(\tau)$ in the path integral.  We can
perform the integral over the value of the constants
 $\Lambda_i$ obtained in these intervals,  with the integral
 running over  positive  or negative reals  respectively for the two
 different kinds of intervals.  To complete the evaluation of the path
 integral, we need to know what amplitude to assign to a crossover
 of $\lambda$ from positive values to negative or the other way
 round.

 This amplitude is not supplied by the action, and so is supplemental
 information needed to make sense of the propagation amplitude.
 It turns out that the freedom of choice here is exactly that
 of a density matrix of the `special' kind  mentioned above
  (exponential of quadratic in
 the field) {\it and} the choice of a `time path'. These are precisely
 the ingredients which define the matrix propagator of the
 real time approach.  The two different signs of the einbein
$\lambda$ on the world line correspond to the two segments of the time path of
the real time formalism, and so furnish the two dimensional
vector space on which the 2x2 matrix propagator acts.

More specifically, we can describe the first quantised
propagator in the following terms.
We consider both the propagation $e^{-i(\boxx+m^2)}$  {\it and} the propagation
$e^{i(\boxx+m^2)}$ on the world
line. We may collect these two possibilities into  a 2x2 matrix form,
getting a world line Hamiltonian ${\rm diag}[(\boxx+m^2), -(\boxx+m^2)]$. Near
the mass shell $\boxx+m^2=0$ the path integral needs regulation. If we add
$-i\epsilon I$ to the Hamiltonian ($I$ is the 2x2 identity matrix) then we
get a diagonal matrix propagator, with \ONETWOP\ and its complex conjugate
as the first and second diagonal entries. But if we regulate instead by
adding $-i\epsilon M$, where $M$ is a non-diagonal matrix, then we get the
matrix propagator for perturbative
calculation of Green's functions in  an `exponential of quadratic' particle
flux.

We compute $M$ for a thermal distribution in Minkowski space;
for the `closed time path'  mentioned above, and for the time path used by
Niemi and Semenoff \ref\NS{A. Niemi and G. Semenoff, Ann. Phys. {\bf 152}
(1981) 105, Nucl. Phys. {\bf B 230} (1984) 181.}.
These two $M$ matrices are not the same, which reflects the fact that the
matrix propagator depends on both $\rho$ and the choice of time path.  For
a curved space example, we consider a $1+1$ spacetime with `sudden'
expansion.  We compute $M$ to obtain the matrix propagator appropriate to the
initial value problem with state the
 `in' vacuum; i.e. for
$\rho={}_{in}|0><0|_{in}$ and time path beginning and ending at past
infinity.

We conclude with a discussion of the significance of our results for
strings, which were the motivation for this study of the first quantised
formalism. The usual first quantised path integrals for strings would give
the analogue of \ONETWOP , which corresponds to  specific boundary
conditions at spacetime infinity. To handle phenomena involving particle
fluxes, we propose extending the world sheet Hamiltonian to a 2x2
Hamiltonian as in the above particle case. Thus we would allow not only
classical deformations of the background field (analogous to the
`exponential of linear' $\rho$ in the particle case) but also particle
flux backgrounds (corresponding to `exponential of quadratic' density
matrices). Studying $\beta$-function equations \ref\SEN{A. Sen, Phys.
Rev. Lett. {\bf 55} (1985) 1846.} for this extended theory should give not
equations between classical fields but equations relating fields and
fluxes. The latter kind of equation, we believe, would be natural to a
description of quantised matter plus gravity.

The plan of this paper is as follows. Section 2 reviews finite
temperature perturbation theory, and discusses the significance of
`exponential of quadratic' density matrices for the curved space theory.
Section 3 translates the finite temperature results of flat space to first
quantised language. Section 4 gives a curved space example.
Section 5 discusses strings. Section 6 is a
summary and  discussion.

\newsec{Propagators in the presence of a particle flux.}

\subsec{Review of the real time formalism.}

The real time method has been extensively developed in
recent years, in a variety of contexts.
A good description of the closed time path and its meaning is given in
\ref\SU{Z-B. Su, L. Yu, and L-Y. Chen in `Thermal Field
Theories' ed. H. Ezawa, T. Arimitsu and Y. Hashimoto (1991) Elsevier
Science Publishers.}, \ref\PAZ{J.P. Paz, in `Thermal Field
Theories' ed. H. Ezawa, T. Arimitsu and Y. Hashimoto (1991) Elsevier
Science Publishers.}.  We give a brief summary below.

Consider a scalar field in Minkowski spacetime (metric signature $+-\dots
-$)
\eqn\TWOONE{S~=~\int dx({1\over 2}
\partial_\mu\phi\partial^\mu\phi-{1\over 2}m^2\phi^2-
{\lambda\over 4!}\phi^4)}
We have the following relation for Green's functions \ref\IZ{C. Itzykson
and J-B Zuber, `Quantum Field Theory'(1980) Mc Graw Hill.}:
\eqn\TWOTWOPP{<0|T\{\phi(x_1)\dots\phi(x_n)\}|0>=
<0|U^{-1}(\infty)T\{\phi_0(x_1)\dots\phi_0(x_n)e^{
-i\int_{t'=-\infty}^{\infty}dt'H_{\rm int}(t')}\} U(-\infty)|0>}
where
\eqn\TWOTHREEPP{U(t)~=~
Te^{-i\int_{t'=-\infty}^{t}dt'H_{\rm int}(t')}}
\eqn\TWOFOURPP{\phi(x)~=~U^{-1}(t)\phi_0(x)U(t)}
and $\phi_0$ is the free field.

{}From \TWOTHREEPP\ we have
\eqn\TWOFIVEPP{U(-\infty)|0>~=~|0>}
If the vacuum is assumed stable:
\eqn\TWOSIXPP{<0|U^{-1}(\infty)~=~e^{i\theta}<0|}
 then we obtain the usual perturbation series
 \eqn\TWOSEVENIPP{<0|T\{\phi(x_1)\dots\phi(x_n)\}|0>~=~
{<0|T\{\phi_0(x_1)\dots\phi_0(x_n)e^{
-i\int_{t'=-\infty}^{\infty}dt'H_{\rm int}(t')}\}|0>\over <0|e^{-i
\int_{t'=-\infty}^{\infty}dt'H_{\rm int}(t')}|0>}}

If the vacuum is {\it not} stable (i.e., \TWOSIXPP\ does
not hold), then we get interaction vertices from $U^{-1}(\infty)$
in \TWOTWOPP . The interaction Hamiltonian acts from
$t=-\infty$ to $t=\infty$ because of the exponential in
 \TWOTWOPP\ and from $t=\infty$ to $t=-\infty$ because
 of  $U^{-1}(\infty)$.  The Green's function can be
 written as
  \eqn\TWOEIGHTPP{<0|T\{\phi(x_1)\dots\phi(x_n)\}|0>~=~
<0|T_C\{\phi_0(x_1)\dots\phi_0(x_n)e^{
-i\int_Cdt'H_{\rm int}(t')}\}|0>}
where $H_{\rm int}$ is integrated along the time path $C$
running from $t=-\infty$ to $t=\infty$ and back to
 $t=-\infty$. $T_C$ represents path ordering along the path $C$.

 We label field operators
on the first part of $C$
  with the subscript $1$  on the second part of $C$
  with the subscript  $2$. Thus for the two point function of the
scalar field there are four possible combinations of subscripts, and the
corresponding propagators are collected into a matrix:
\eqn\TWOTENPP{D(x_2,x_1)~=~\pmatrix{{}_{in}
<0|T[\phi(x_2)\phi(x_1)]|0>_{in} & {}_{in}<0|\phi(x_1)\phi(x_2)|0>_{in}\cr
{}_{in} <0|\phi(x_2)\phi(x_1)|0>_{in}& {}_{in}<0|\tilde
T[\phi(x_2)\phi(x_1)]|0>_{in}\cr}}
where we have added the subscript `in' to the vacuum to emphasize
that it is the vacuum state at $t=-\infty$. ($\tilde T$ describes anti
time ordering.)

Perturbation in the coupling $\lambda$ is described by the following
diagram rules \ref\LANDS{N.P. Landsman and Ch.G.
Van Weert, Phys. Rep. {\bf 145} (1987) 141.}. The external insertions are all
of type $1$.
(This can be seen from the way we arrived at \TWOEIGHTPP\
{}from \TWOTWOPP .)  The vertices
{}from the perturbation term must have either all legs of type $1$ or all
legs of type $2$. Attach a factor $(-i\lambda)$ to the vertex in the
former case, $(i\lambda)$ in the latter. The propagator is given by the
matrix \TWOTENPP\ and can connect vertices of type $1$ to vertices of type
$2$. Summing all Feynman diagrams with these rules
gives the correlation function  for the initial value problem with initial
state the vacuum at $t=-\infty$.

  As a simple example  where the second leg of the time path
  makes a difference,   consider the free scalar field in
$0+1$ spacetime dimensions.  Let the only perturbation be
the time dependent term $\mu\int dt
\delta(t-t_0):\phi^2(t):$. We compute  to first order in $\mu$
the two point correlator for field insertions at $t_1<t_0<t_2$,
using time path $C$:
\eqn\TWOEIGHT{\eqalign{<\phi(t_2)\phi(t_1)>~=~&
2\mu[<0|\phi(t_2)\phi(t_0)|0> <0|\phi(t_0) \phi(t_1)|0>\cr &\quad
{}~+~<0|\phi(t_0)\phi(t_2)|0> <0|\phi(t_0)\phi(t_1) |0>]\cr}}
Without the second leg of the time path, we just get the first term
on the RHS.  The latter answer gives
the amplitude to start in the  interaction picture vacuum at $t=-\infty$ and to
also end in interaction picture vacuum at $t=\infty$. The  missing  term
corresponds
to  correcting the state at
$t=\infty$ away from the vacuum $|0>$, to be such that it evolves from
$|0>$ after suffering the perturbation at $t_0$. Thus the complete
answer \TWOEIGHT\ gives a Heisenberg picture expectation value,
and is thus the only physically meaningful object to be computed
in this situation.

 We can rewrite \TWOEIGHTPP\ as
   \eqn\TWONINEPP{<0|T\{\phi(x_1)\dots\phi(x_n)\}|0>~=~
\Tr\{\rho_0T_C\{\phi_0(x_1)\dots\phi_0(x_n)e^{
-i\int_Cdt'H_{\rm int}(t')}\}\}/\Tr\rho_0}
with $\rho=|0><0|$ the density matrix of a  pure state.
But we can immediately extend the above relation to
to the case where $\rho_0$ is replaced by an arbitrary density matrix
$\rho$, not necessarily representing a pure state, and not
necessarily describing a distribution in equilibrium.  To develop
diagrammatic perturbation theory, however, we need to have the
Wick decomposition for expectation of operator products.  This
is achieved for a special class of density matrices, which
have the form `exponential of quadratic in the field operator $\phi$'.
Departures of $\rho$ from this special form are handled
perturbatively  just like interaction vertices.  We discuss the
significance of this special class of $\rho$ in the next subsection.

As an example of mixed
state density matrices, let   $\rho$ at $t=-\infty$ be of thermal
form with temperature $(\beta)^{-1}$. Then the matrix propagator is
\eqn\TWOSEVENPPP{D(p)~=~\pmatrix{{i\over
p^2-m^2+i\epsilon}+2\pi n(p)\delta(p^2-m^2) &
2\pi[(n(p)+\theta(-p_0)]\delta (p^2-m^2) \cr 2\pi[n(p)+\theta(p_0)]
\delta(p^2-m^2) & {-i\over p^2-m^2-i\epsilon}+2\pi n(p)\delta(p^2-m^2)\cr}}
where
\eqn\TWOTHREEP{n(p)~=~(e^{\beta|p_0|}-1)^{-1}}
is the number density of particles for the free scalar field at
temperature $(\beta)^{-1}$.
 With this change of matrix
 propagator, perturbation theory has the same rules as  for the vacuum
 case.

 Another approach leading to a  similar diagrammatic structure was given
 by  Niemi and Semenoff \NS . In the periodic
 imaginary time description of temperature, we evolve
 the
field theory in imaginary time from $t=0$ to $t=-i\beta$, and identify
these two time slices. In the `real time' approach of Niemi and Semenoff
 we use instead the following path $C_1$ in complex $t$ space  to
connect these two slices
\eqn\TWOTWO{\eqalign{C_1:\quad\quad&I:\quad -\infty\rightarrow\infty\cr
&II:\quad\infty\rightarrow \infty-i\beta/2\cr
&III:\quad\infty-i\beta/2\rightarrow -\infty-i\beta/2\cr &IV:\quad
-\infty-i\beta/2\rightarrow -\infty-i\beta \cr}} The concept of time
ordered correlation functions is  replaced by `path ordered'
correlation functions, with $t$ running along the above path $C_1$. One
argues that the parts $II$, $IV$ of $C_1$ can be ignored. Field operators
on part $I$ are labelled with the subscript $1$ while on part $III$ are
labelled with the subscript $2$.
The matrix propagator becomes
\eqn\TWOTHREE{D_{NS}(p)~=~\pmatrix{{i\over
p^2-m^2+i\epsilon}+2\pi n(p)\delta(p^2-m^2) &
2\pi[(n(p)(n(p)+1)]^{1/2}\delta (p^2-m^2) \cr 2\pi[n(p)(n(p)+1)]^{1/2}
\delta(p^2-m^2) &
{-i\over p^2-m^2-i\epsilon}+2\pi n(p)\delta(p^2-m^2)\cr }}

Perturbation theory based on the contours $C$ and $C_1$ are not
equivalent, even for zero temperature \PAZ .  The propagators  \TWOSEVENPPP\
and
\TWOTHREE\  are not the same, for $\beta\rightarrow\infty$.
Thus the natrix
propagator depends both on the particle distribution
and on the time path.

Note that the temperature dependent terms in \TWOSEVENPPP\ or
\TWOTHREE\  are all on shell. This is a manifestation of the
fact that these terms describe a flux of real particles.  The
expression $\delta(p^2-m^2)$ does not have a Euclidean counterpart.  (In
Euclidean space one would consider $(p^2+m^2)$ which is positive
definite.) Thus temperature is a phenomenon of Minkowski signature
spacetime, where Green's functions are subject to the addition of
solutions of the homogeneous field equation.

To understand the origin of the on shell terms, let us consider the
element $D_{11}$ in \TWOTHREE . We can decompose the free scalar field in
Minkowski space into Fourier modes. Each quantised mode is a harmonic
oscillator with frequency $\omega({\bf p})=({\bf p}^2+m^2)^{1/2}$. The two
point function $<T[\phi(x_2)\phi(x_1)]>$ reduces, for each Fourier mode,
to $<T[\hat q(t_2)\hat q(t_1)]>$ for each harmonic oscillator. Thus we
focus on a single oscillator (we suppress its momentum label). Let
$t_2>t_1$. Then
\eqn\TWONINE{\eqalign{<\hat q(t_2)\hat q(t_1)>~\equiv~&
\sum_n e^{-\beta(n+{1\over 2})
\omega}<n|T[\hat q(t_2)\hat q(t_1)]|n>/\sum_n e^{-\beta(n+{1\over 2})
\omega} \cr
&={<n+1>\over 2\omega}e^{-i\omega(t_2-t_1)}~+~{<n>\over 2\omega}
e^{i\omega (t_2-t_1)} \cr}} The first term on the RHS corresponds to the
stimulated emission of a quantum at $t_1$ with absorption at $t_2$. The
second term corresponds to the annihilation at time $t_1$ of one of the
existing quanta in the thermal bath, and the subsequent transport of a
hole from $t_1$ to $t_2$, where another particle is emitted to replace the
one absorbed from the bath.  The Fourier transform in time of \TWONINE\ is
\eqn\TWOTEN{\int_{-\infty}^{\infty}dt
e^{-i\omega t}<T[\hat q(t_2)\hat q(t_1)]> ~=~ {i\over
\omega'^2-\omega^2+i\epsilon} ~+~2\pi<n>\delta(\omega'^2-\omega^2 )} which
gives the matrix element $D_{11}$ in \TWOTHREE .

Thus the correction to $<T[\phi(x_2)\phi(x_1)]>=D_{11}(x_2,x_1)$ (and
other elements of $D$) due to the particle flux is not an effect of
interactions with the particles of the bath. This correction arises from
the possible exchange of the propagating particle with identical real
particles in the ambient flux. Using such a corrected propagator with the
interaction vertices gives the interaction of the bath particles with the
propagating particle.  With this understanding of the propagator we see
that the real-time formalism is really a  covariant concept.
In a general spacetime, single particle wavefunctions are just solutions
of the free field equation. If there is more than one particle populating
the same wavefunction, then we must keep track of `exchange
terms' in computing correlation functions. The matrix propagator
does this bookkeeping; the terms with $\delta(p^2-m^2)$ in the above
would be replaced with  the
appropriate solutions of the wave equation in the spacetime.

\subsec{`Exponential of quadratic' density matrices.}

When doing perturbation theory on flat space, we usually assume that the vacuum
is stable: $|0>_{out}=|0>_{in}$ (upto a phase). As we
saw above, when we do not have this stability then we
have to use a 2x2 matrix propagator. Stability may be lost
either because the spacetime geometry is time-dependent, or
because there exists a distribution of particles in the initial state.
With gravity we would typically encounter both these causes:
the time dependent geometry  creates particle fluxes which
evolve on (and  modify) the geometry.  Thus it becomes natural
to construct the theory with arbitrary initial state and to not consider
a `vacuum theory' at all.

In perturbation theory we deal easily with a special  class
of states: the coherent states.  These have the form $e^{A[\phi]}|0>$
where $A$ is an operator linear in the field variable $\phi$.  The
operator $A$ generates a change in the classical value of the field;
shifting  the field variable by this classical value removes the
effect of $A$ from Feynman diagrams.  In studying first
quantised propagation about a background, this shift has already
been made (to zeroth order in the coupling).

The discussion of the real time method revealed that there
is another kind of field configuration that is also easy to deal with.
This is the class of configurations expressed as density matrices
$\rho=e^{B[\phi]}$, where $B$ is quadratic in the field operator.
The special role of these `exponential of quadratic' density
matrices
  is due to the fact that Wick's theorem extends to
correlators computed with such $\rho$
\ref\HENNING{P.A. Henning, Nucl. Phys. {\bf B 337} (1990) 547.}. Thus for
operators $A_i$ linear in the field,
\eqn\TWOTHREETWO{<A_1\dots A_n>~\equiv~
{1\over \Tr \rho} \Tr \{ \rho
A_1\dots A_n \}~=~\sum_{\rm permutations}<A_{i_1}A_{i_2}>\dots <A_{i_{n-1}}
A_{i_n}>} We sketch a proof of \TWOTHREETWO\ in the appendix.

The physical importance of `exponential of quadratic' density matrices
is seen from the different instances where they occur:

a) \quad  The thermal distribution has such a density matrix:
\eqn\TWOONEPP{\rho_{\rm thermal}~=~\prod_{{\rm modes}~i}
e^{-\beta \omega_i \ad_i a_i}}

b) \quad  In time dependent geometries one considers the
Bogoliubov transformation relating different expansions of the
field variable (we supress the mode index):
\eqn\TWOTWOPP{\eqalign{\phi~=&~a_{in}f_{in}~+~\ad_{in}f_{in}^*\cr
\phi~=&~a_{out}f_{out}~+~\ad_{out}f_{out}^* \cr}}
The `in' and `out' vacuua are given by
\eqn\TWOTHREEPP{a_{in}|0>_{in}~=~0, \quad\quad a_{out}
|0>_{out}~=~0.}
The `in' vacuum can be expressed as
\eqn\TWOFOURPP{|0>_{in}~=~e^{b\ad_{out}\ad_{out}}|0>_{out} }
so that it contains a flux
of `out' particles.
If the initial state is $|0>_{in}$ then the density matrix for the initial
value problem is $\rho_0=|0>_{in}{}_{in}<0|$.  $\rho_0$
 can be expressed as
\eqn\TWOFIVEPP{\eqalign{
\rho_0~=&~e^{b\ad_{out}\ad_{out}}|0>_{out}{}_{out}<0|e^
{b^*a_{out}a_{out}}\cr
=&{\rm lim}_{\beta\rightarrow\infty}[\rho/\Tr\rho]\cr}}
where
\eqn\TWOSIXPP{\rho~=~ e^{b \ad_{out}\ad_{out}} e^{-\beta \ad_{out}
a_{out}} e^{b^* a_{out} a_{out}} }
and the limit is established by inserting  complete set of
`out' number operator eigenstates between the exponentials in
\TWOFIVEPP .

Thus the `exponential of quadratic' density matrices
 \eqn\TWOSEVENPP {\rho ~=~\prod_i
 e^{\gamma a_i a_i }e^{\alpha a_i^\dagger
 a_i^\dagger}e^{-\beta a_i^\dagger a_i}}
 unify the treatment of thermal fluxes and the fluxes created by
 gravitational fields.
An ensemble like \TWOONEPP\  might describe the distribution at the initial
state of the Universe. In fact the notion of `thermal' distributions
 has to be extended to the class \TWOSEVENPP\ to be useful
 in curved space.  Suppose we choose a time co-ordinate $t$ and start with
a  distribution $e^{-\beta H}$, thermal for  the Hamiltonian giving
evolution in $t$. As the Universe expands, the distribution will not
remain thermal, in general. Redshifting of wavelengths gives an obvious
departure from thermal form if the field has a mass or is not conformally
coupled. But even  a massless conformally coupled field departs from
thermal form if the time co-ordinate $t$ is not appropriately chosen
\ref\SEMEW{G. Semenoff and N. Weiss, Phys. Rev. {\bf D 31} (1985) 689.}.
However,  the density matrix remains within the class
\TWOSEVENPP , if it starts in this class, even for a massive field.

 The Bogoliubov transformation is important in
obtaining the stress energy of particles created in the gravitational field.
For example in the black hole geometry the state is the Kruskal vacuum.
The Bogoliubov transform to the Schwarzschild modes allows us  to
see readily  the  $<T_{\mu\nu}>$ of emitted radiation at spatial infinity.
The class \TWOSEVENPP\ is closed under Bogoliubov transformations
relating different field expansions like \TWOTWOPP .  It is the
natural class to arise in a covariant formulation where the density
matrix $\rho_0$ discribed above is one of the allowed initial state
specifications.

How do we compute the backreaction on the
gravitational field from particles described by
\TWOSEVENPP ? For a coherent state, the stress tensor is
given, ignoring quantum corrections, by the classical
value of $<T_{\mu\nu}>$ computed for the classical field value
implied in the coherent state. But the density matrices \TWOSEVENPP\
do not approximate a classical configuration; indeed
the   dispersion in $T_{\mu\nu}$ in each mode is
always of the order of   $T_{\mu\nu}$ itself \ref\FORD{L.H. Ford and
C-I Kuo, Preprint TUTP-93-1 (gr-qc 9304008) (1993)}.  Can we
take into account the backreaction from such distributions? By
incorporating the `exponential of quadratic' density matrices in the
propagator the real time formalism allows us to do precisely that.
Thus even though such configurations are far from classical
the Feynman diagram technique can handle them, provided we
make  the
extension to the 2x2 matrix propagator.

What if $\rho$ was not well approximated by the form
$e^{A[\phi]+B[\phi]}$? No easy route appears available
for studying backreaction in that case. {\it Small} deviations
{}from the above form can be handled perturbatively as
`correlation kernels', which behave as extra, nonlocal vertices
for the perturbation series. For a description of the expansion in
 correlation kernels and interaction vertices, see \HENNING .

 Thus in the real time method we are relying on the assumption
 that for the situations of interest the state is well approximated by
 coherent and `exponential of quadratic' parts. From the discussion
 above about  particle creation, we conclude that allowing coherent
 states alone will not be adequate for a theory of gravity,  particularly
  in regions of strongly varying garvitational  field.  But if we
  set all  couplings (other than gravity) to be small,  we should
  be able to validate the above assumption about $\rho$, and carry
  out real time perturbation theory.

We should distinguish two different limits in which the physics of fluxes
may be studied. One limit is where the collisions are so rapid that
approximate thermal equilibrium is maintained at all times, and we need
only let $\beta$ be a function of time. The other limit is that of kinetic
theory, where we assume that collisions are rare; particle wavefunctions
evolve on the time-dependent background, and collisions between these
particles are taken into account by  perturbation theory. Our approach
assumes the latter limit.

\newsec{The first quantised formalism.}

\subsec{The regulator matrix.}

The Feynman propagator for a scalar field in Minkowski space can be
written as
\eqn\THREEONE{D_F(p)~=~{i\over p^2-m^2+i\epsilon}~=~\int_{\tilde
\lambda=0}^\infty d \tilde \lambda e^{i\tilde\lambda (p^2-m^2+i\epsilon)}}
\THREEONE\ can be used to express $G_F(p)$ in a first quantised language,
with $p^2=-\boxx$. (See for example \ref\PARISI{G. Parisi, `Statistical
Field Theory' (1988) Addison Wesley.}, \ref\KAKU{M. Kaku, `Introduction to
Superstrings' (1988) Springer-Verlag.}.) The Hamiltonian on the world line
is $\boxx+m^2$, evolution takes place in a fictitious time for a duration
$\tilde \lambda$, and this length $\tilde \lambda$ of the world line is
summed over all values from $0$ to $\infty$.

 Let us write the matrix propagator \TWOTHREE\ in a similar fashion
\eqn\THREETWO{D_{NS}(p)~=~ \int_0^\infty d\tilde \lambda
e^{-i\tilde\lambda H-\epsilon\tilde\lambda M}} where
\eqn\THREETWOI{H=\pmatrix
 {-(p^2-m^2)&0\cr 0&(p^2-m^2)\cr }, \quad\quad M=
\pmatrix{1+2n(p)&-2\sqrt{n(p)(n(p)+1)}\cr -2\sqrt{n(p)(n(p)+1)}&
1+2n(p)\cr}} ($n(p)$ is given by \TWOTHREEP .) This matrix world line
Hamiltonian has the following structure. If we forget the term multiplying
$\epsilon$, then the $a=1$ component of the vector state on the world line
evolves as $e^{i \tilde \lambda (p^2-m^2)}$ while the $a=2$ component
evolves as $e^{-i\tilde\lambda(p^2-m^2)}$. To define the path integral we
need the regulation from $\epsilon$ at the mass shell $p^2-m^2=0$. But the
regulator matrix is not diagonal, for nonzero temperature. Thus
transitions are allowed from state $1$ to state $2$. As we shall show below, in
the limit
$\epsilon\rightarrow 0^+$, which we must finally take, these transitions
occur only on the mass shell. (The absolute values of the entries in $M$
are not significant, because $\epsilon$ goes to $0$, but the relative
values are.)

Similarily we can write the matrix propagator \TWOSEVENPPP\ in the form
\THREETWO\ with $H$ as in \THREETWOI\ but
\eqn\THREETHREE{M~=~\pmatrix{1+2n(p)&-2\sqrt{n(p)(n(p)+1)}e^{-\beta
p_0/2}\cr -2\sqrt{n(p)(n(p)+1)}e^{\beta p_0/2} & 1+2n(p)\cr}} Again we
have the  $a=1,2$ states propagating on the world line with Hamiltonians
$\mp(p^2-m^2)$. Transitions between $a=1,2$ are again of order $\epsilon$,
but are different from those in \THREETHREE .

Looking at these examples the following picture emerges. To obtain the
matrix propagator in a many-body situation (with `exponential of
quadratic' density matrices) we need to consider both evolutions
$e^{-i\tilde\lambda H}$ and $e^{i\tilde \lambda H}$ on the world line. A
damping  factor is needed to define the first quantised path integral,
near the mass shell. But the regulator matrix $M$ need not be diagonal.
 The off-diagonal terms of $M$ encode the strength
 of an `exponential of quadratic flux'.  These terms also
 depend on the choice of the time path, as we see from the difference in $M$
for the contours $C$ and $C_1$.

 The limit $\epsilon\rightarrow 0$ implies that the effect of the regulator
matrix $M$ is felt only on-shell (i.e. for $p^2-m^2=0$).  Equivalently, we
may say that only world lines of infinite length ($\tilde
\lambda =\infty$) see the
regulator matrix. To see this, let $M$ and $M'$ be two different regulator
matrices. We write
\eqn\FIVEONE{\eqalign{D_{M'}~=~&{\rm lim}_{L\rightarrow\infty}
{\rm lim}_{\epsilon\rightarrow 0^+}\{[\int_0^L d\tilde\lambda
e^{-iH\tilde\lambda-\epsilon M \tilde\lambda}+\int_L^\infty d\tilde\lambda
e^{-iH\tilde\lambda-\epsilon M \tilde\lambda}] \cr &~+~[\int_L^\infty
d\tilde\lambda e^{-iH\tilde\lambda-\epsilon M'
\tilde\lambda}-\int_L^\infty d\tilde\lambda e^{-iH\tilde\lambda-\epsilon M
\tilde\lambda}]\cr &~+~[\int_0^L d\tilde\lambda
e^{-iH\tilde\lambda-\epsilon M' \tilde\lambda}-\int_0^L d\tilde\lambda
e^{-iH\tilde\lambda-\epsilon M \tilde\lambda}] \}\cr}} The first square
bracket on the RHS is $D_M$, the last vanishes with the indicated limits,
while the second has support only on world lines of infinite length.
Let us take $M=I$, the identity matrix. Then $D_M={\rm diag}\{
{i\over p^2-m^2+i\epsilon},{-i\over P^2-m^2-i\epsilon}\}$.  Thus
other propagators (obtained by using $M$ not the identity matrix)
differ from this basic one only by the contribution of world lines
of infinite length.

\subsec{Quantising the relativistic particle.}

What is the origin of the two components $a=1,2$ of the state on the world
line? We would like to offer the following heuristic `derivation' as a
more physical description  of the matrix structure in \THREETWO .

The geometric action for a scalar particle is
\eqn\THREEFOUR{S~=~\int_{X_i}^{X_f} mds~=~\int_{X_i}^{X_f} m(X^\mu,_\tau
X_\mu,_\tau )^{1/2}d\tau~\equiv~\int L d\tau} where $\tau$ is an arbitrary
parametrisation of the world line. The canonical momenta
\eqn\THREEFIVE{P_\mu~=~{\partial L\over \partial X^\mu,_\tau}~=~{mX_\mu,_
\tau\over (X^\mu,_\tau X_\mu ,_\tau)^{1/2}}}
satisfy the constraints
\eqn\THREEFIVE{P^\mu P_\mu-m^2~=~0}
We choose the range of the parameter $\tau$ as $[0,1]$. Following the
approach in \KAKU , we impose the constraint at each $\tau$ through a
$\delta$-function:
\eqn\THREESEVEN{\delta(p^2(\tau)-m^2)~=~{1\over 4\pi}\int_{-\infty}^\infty
d \lambda(\tau)e^{-i\lambda/2(p^2(\tau)-m^2)}} The path integral amplitude
to propagate from $X_i$  to $X_f$ becomes
\eqn\THREEEIGHT{G(X_2,X_1)~\equiv~
N\int{D[X]D[P]D[\lambda]\over{\rm Vol}[{\rm Diff}]}
e^{i\int_0^1 d\tau[P_\mu(\tau)
X^\mu,_\tau(\tau)-\lambda/2(\tau)(p^2(\tau)-m^2)]}} where $N$ is a
normalisation constant, $P_\mu X^\mu ,_{\tau}=m(X^\mu,_\tau X_\mu
,_\tau)^{1/2}$ is the original action \THREEFOUR\ and we have divided by
the volume of the symmetry group, which which is related to
$\tau$-diffeomorphisms in the manner discussed below. (The
$\delta$-function constraint on the momenta and dividing by ${\rm
Vol}[{\rm Diff}]$ remove the two phase space co-ordinates redundant in the
description of the particle path.)

There are two ways to consider the symmetry of the action
\THREEEIGHT .
The action is invariant under
\eqn\THREENINE{\eqalign{{\cal S}_1:\quad\quad &\delta
X^\mu(\tau)~=~h(\tau) P^\mu(\tau) \cr &\delta P_\mu(\tau)~=~0 \cr &\delta
\lambda (\tau)~=~ h(\tau),_\tau \cr}} and
\eqn\THREETEN{\eqalign{{\cal S}_2: \quad\quad&\delta
X^\mu (\tau)~=~\epsilon(\tau)\lambda(\tau) P^\mu(\tau) \cr &\delta
P_\mu(\tau)~=~0 \cr &\delta
\lambda(\tau)~=~(\epsilon(\tau)\lambda(\tau)),_\tau \cr}} The difference
between ${\cal S}_1$ and ${\cal S}_2$ is best seen by considering the
finite transformations on $\lambda$:
\eqn\THREEELEVEN{{\cal
S}_1:\quad\quad\lambda'(\tau)~=~\lambda(\tau)~+~{dh(\tau)\over d\tau}}
\eqn\THREETWELVE{{\cal S}_2:\quad\quad \lambda'(\tau)~=~
{d\tau \over d\tau'}(\tau) \lambda(\tau)}
Using ${\cal S}_1$ we can gauge
fix any function $\lambda(\tau)$ to any other function $\lambda_1(\tau)$,
provided $\lambda$, $\lambda_1$ have the same value of
\eqn\THREETHIRTEEN{\int_0^1\lambda(\tau)d\tau~\equiv~\Lambda}
With such gauge fixing we obtain not
a Green's function but a solution of the homogeneous Klein-Gordon
equation.  Thus the Fourier transform of  \THREEEIGHT\
gives
\eqn\THREEIONE{\tilde G(p)~=~\delta(p^2-m^2)}
for a suitable normalisation of the measure.

With ${\cal S}_2$, $\lambda$ transforms as an einbein under the
diffeomorphism $\tau\rightarrow \tau '(\tau)$. Note that for regular
$\epsilon(\tau)$, $\lambda$ either changes sign for no $\tau$ or for all
$\tau$. We take ${\rm Diff}$ as the group of regular diffeomorphisms
connected to the identity; then $\lambda$ does not change sign under
the allowed diffeomorphisms.   These
diffeomorphisms cannot gauge-fix $\lambda(\tau)$ to any preassigned
function $\lambda_1(\tau)$. We have the obvious  restriction coming
{}from
\THREETHIRTEEN , where $\Lambda$ may now be interpreted as the length of
the world line. This restriction is usually assumed to mean that the
length of the world line is the only remaining parameter after
gauge-fixing. But what we find instead is that there is a
further complication: there is a discrete infinity of
classes, each with one or more continuous parameters.  One
class comes from configurations $\lambda(\tau)$ which are everywhere
positive. This class can be gauge-fixed  with the
allowed diffeomorphisms to have
\eqn\THREEI{\dot\lambda(\tau)~=~0, \quad\quad \int_0^1 d\tau \lambda
(\tau)~=~\Lambda } with $0<\Lambda<\infty$. Similarily, the set of
everywhere negative $\lambda(\tau)$ can be gauge-fixed as in \THREEI\ but
with $-\infty<\Lambda<0$.
Thus these classes give for the Fourier transform of \THREEEIGHT\
\eqn\THREEITWO{\eqalign{G_{\Lambda>0}(p)~=&~
{1\over 4\pi}\int_{0}^\infty
d \lambda(\tau)e^{-i\lambda/2(p^2(\tau)-m^2)}~=~
{i\over p^2-m^2+i\epsilon},\cr
G_{\Lambda<0}(p)~=&~
{1\over 4\pi}\int_{-\infty}^0
d \lambda(\tau)e^{-i\lambda/2(p^2(\tau)-m^2)}~=~
{-i\over p^2-m^2-i\epsilon}\cr}}
respectively. (We have added the $\epsilon$ term for
convergence to each sector; this term was not in the action.)
 Keeping the first class alone gives the Feynman
propagator for particles, while the second gives its complex
conjugate.\foot{A restriction to the range  $(0,\infty)$ for $\Lambda$ can
be naturally obtained using a Newton-Wigner formalism \ref\HAR{J.B.
Hartle and K.V. Kuchar, Phys. Rev. {\bf D 34} (1986) 2323.}. Here the
particle travels only forwards in the time co-ordinate $X^0$, thus it is
not a co-variant approach.} Note
that these two objects are the diagonal entries of  \TWOTHREE\ for
$\beta=\infty$.

But in the path integral over $\lambda$ in \THREEEIGHT\ we also
have the class of  $\lambda(\tau)$ which are positive for $0<\tau<\tau_1$,
negative for
$\tau_1<\tau<1$. The group of orientation preserving diffeomorphisms can
gauge fix this to
\eqn\THREEII{\dot\lambda(\tau)=0 ~~{\rm for} ~~\tau\ne\tau_1,
{}~~~~\int_0^{\tau_1} d\tau \lambda(\tau)=\Lambda_1, ~~\int_{\tau_1}^1
d\tau
\lambda(\tau)=\Lambda_2}
with $0<\Lambda_1<\infty$, $-\infty<\Lambda_2<0$. We would like to
identify this sector as the contribution to the amplitude to start with a
state of type $1$ and end with a state of type $2$ (the off-diagonal
element $D_{12}$ of the matrix propagator).  A
general sector has a given number of alternations
in the sign of $\lambda$, and in each interval of
constant sign we gauge fix $\lambda$
to a constant $\Lambda_i$.  We can perform the integration
over the variables $\Lambda_i$ appearing in each sector,  but
we should also specify a `transition amplitude' for each point $\tau_i$
where $\lambda$ changes sign.  We allow this
 amplitude to depend on the
states on both sides  of $\tau_i$, and also on whether
$\lambda$ changes from positive to negative or vice versa.
 This amlplitude is not supplied by the original action;
it is supplementary information needed for determing the propagator.

We would like to identify  $\lambda>0$ with  particles of the type 1
field  and $\lambda<0$  with  particles of the
type two field in the language of section 2.  To obtain the matrix
propagator element $D_{ji}$  we need to  add together all sectors
for $\lambda(\tau)$ beginning as type
$i$ and ending as type $j$.
Take the example \THREETWO , \THREETWOI\  and write
\eqn\THREEIPONE{D(p)~=~~=~\int_0^\infty d\tilde \lambda
e^{-i\tilde\lambda H-\epsilon\tilde\lambda M} ~=~ \int_0^\infty d\tilde \lambda
e^{-i\tilde\lambda H'-\epsilon\tilde\lambda M'}}
\eqn\THREEIPTWO{\eqalign{H'=&\pmatrix
 {-[p^2-m^2+i\epsilon(1+2n(p))]&0\cr 0&[p^2-m^2-i\epsilon(1+2n(p))]\cr },\cr
M'=&
\pmatrix{0&-2\sqrt{n(p)(n(p)+1)}\cr -2\sqrt{n(p)(n(p)+1)}&
0\cr}\cr}}

Expanding \THREEIPONE\ in a perturbation series in $M'$,
\eqn\THREEIPTHREE{\eqalign{D(p)~=&~\int_0^\infty d\lambda_1
e^{-iH'\lambda_1}~+~ \int_0^\infty d\lambda_1
d\lambda_2 e^{-iH'\lambda_1}(-\epsilon M') e^{-iH'\lambda_2}~+~
\dots\cr
=&~D^0~+~D^0(-\epsilon M')D^0~+~\dots\cr}}
where
\eqn\THREEIPPONE{D^0~=~\pmatrix{{i\over p^2-m^2+i\tilde\epsilon}&0\cr
0&{-i\over p^2-m^2-i\tilde\epsilon}\cr}}

$D_{12}$ for example gets a contribution from every term with an odd
number of $M'$ insertions:
\eqn\THREEIPFOUR{\eqalign{D_{12}(p)~=&~-\epsilon D^0_{11}M'_{12}D^0_{22}
{}~-~\epsilon^3 D^0_{11}M'_{12}D^0_{22}M'_{21}D^0_{11}M'_{12}D^0_{22}~+~
\dots \cr
=&~\sum_{m=0}^\infty ({1\over (p^2-m^2+\tilde \epsilon^2})^m
(2\tilde\epsilon{\sqrt{n(p)(n(p)+1)}\over 1+2n(p)})^{2m-1}\cr
=&~2\pi\delta(p^2-m^2)\sqrt{n(p)(n(p)+1)}\cr}}
where $\tilde\epsilon=\epsilon(1+2n(p))$.

 Thus suppose the propagators for $\lambda>0$, $\lambda<0$ are
 ${\pm i\over p^2-m^2\pm i\tilde\epsilon}$ respectively.  Let us
 choose the amplitude for $\lambda$ to change sign
 by requiring that momentum $p$ goes to $p$, and the
 associated amplitude is $2\tilde\epsilon{\sqrt{n(p)(n(p)+1)}\over 1+2n(p)}
 =\tilde\epsilon{\rm sech}(\beta|p_0|/2)$.  Then summing all the sectors that
contribute to each $D_{ji}$ we
 get the matrix propagator \TWOTHREE .

 To obtain \TWOSEVENPPP , we must take the amplitude for $\lambda$ to
 go from positive to negative as $\tilde\epsilon {\rm sech}(\beta|p_0|/2)
e^{\beta p_0/2}$, and for negative to positive as
 $\tilde\epsilon {\rm sech}(\beta|p_0|/2)e^{-\beta p_0/2}$.

{\it Note:} \quad To achieve the interpretation of a matrix of propagators
as arising from the above `sum over sectors', the matrix itself
must be very special.  From the symmetry of \TWOTHREE\ we
find that $M$ must have the form $\pmatrix{a&b\cr b&a}$.
Because of the limit $\epsilon\rightarrow\infty$ only the ratio
$b/a$ affects the obtained propagator. As we saw above, this
ratio (which directly gave the amplitude for reversing world line orientation)
got determined by the value of $D_{12}$. But now there are no
free parameters to adjust for obtaining $D_{11}$. A direct calculation
gives
\eqn\THREEIPPTWO{\eqalign{D_{11}(p)~=&~
D^0_{11}~+~\epsilon^2D^0_{11}M'_{12}D^0_{22}M'_{21}D^0_{11}~+~
\dots \cr
=&~\sum_{m=0}^\infty {i(2\tilde \epsilon {\sqrt{n(p)(n(p)+1)}\over
1+2n(p)})^{2m-1})^{2m}\over (p^2-m^2+i\tilde\epsilon)
((p^2-m^2)^2+\tilde\epsilon ^2)^m}\cr
=&~{i\over p^2-m^2+i\tilde\epsilon}~+~2\pi\delta(p^2-m^2)(1+2n(p))\cr}}
which agrees with $D_{11}$ in \TWOTHREE . Thus the matrix of propagators
obtained from the real time formalism has, at least for this
example, the special form required for the first quantised interpretation to
work. \TWOSEVENPPP\ also has the required form. $M$ has
unequal off diagonal terms but their product remains the same as
in the previous example; thus we get the same $D_{11}$, in
agreement with \TWOSEVENPPP .

Thus we have verified in these examples that the amplitude
for einbein sign change reflects the presence of particle flux in the first
quantised formalism. In section  4 we will
present an example where the flux created by spacetime
expansion is handled in  a similar way.

\subsec{BRST formalism.}

One might wonder if the more formal BRST quantisation of the relativistic
particle would resolve the above issues about different possible
quantisations. We follow the notation in  \GOVAERTS .  We introduce the
canonical conjugate $\pi$ for $\lambda$
($[\lambda,\pi]=i$) and ghosts $(\eta^1,{\cal P}_1)$, $(\eta^2,{\cal
P}_2)$ ($[\eta^i,{\cal P}_j]=-i\delta^i_j$) for the two constraints
$\pi=0$ and ${1\over 2} (P^2-m^2)=0$ respectively. The BRST charge
\eqn\THREETHREEONE{Q~=~\eta^1\pi~+~{\eta^2\over 2}(P^2-m^2)}
is nilpotent ($Q^2=0$), and gives the variations:
\eqn\THREETHREETWO{\eqalign{
\delta X^\mu=-i[X^\mu,Q]=\eta^2P^\mu\quad\quad &\delta P_\mu=0
\cr
\delta \lambda =\eta^1 \qquad\qquad\qquad&\delta \pi=0 \cr
\delta \eta^1=0 \qquad\qquad\qquad&\delta {\cal P}_1=-\pi  \sim 0 \cr
\delta \eta^2=0 \qquad\qquad\qquad
&\delta {\cal P}_2=-{1\over 2}(P^2-m^2) \sim 0 \cr }} The equation of
motion (obtained after gauge fixing) gives $\dot
\eta^2=\eta^1$, which agrees with \THREENINE .

But we can define another  nilpotent BRST charge
\eqn\THREETHREETHREE{Q'~=~{\eta^1}'\pi'\lambda'~+~
{1\over 2}\eta'^2(P^2-m^2)} which generates the symmetry
\eqn\THREETHREEFOUR{\eqalign{
\delta {X^\mu}'={\eta^2}'{\lambda}'{P^\mu}'\quad\quad\quad\quad &\delta
{P_\mu}' =0
\cr
\delta {\lambda}' ={\lambda}'{\eta^1}' \qquad\qquad\qquad
&\delta {\pi}'=-{\eta^1}'{\pi}'\sim 0 \cr
\delta {\eta^1}'=0 \qquad\qquad\qquad
&\delta {\cal P}_1'=-{\lambda}'{\pi}'  \sim 0 \cr
\delta {\eta^2}'=0 \qquad\qquad\qquad&\delta {\cal P}_2'=-{1\over 2}
({P^2}'-m^2) \sim 0 \cr }} The equation of motion gives
${\dot\eta}^{2\prime}={\eta^1}'{\lambda}'$, which suggests that we should
identify the above symmetry with ${\cal S}_2$.

The symmetries $Q$ and $Q'$ are related through the identifications
\eqn\THREETHREEFIVE{\pi={\pi}'{\lambda}', \quad\quad\quad\quad
\lambda=log{\lambda}'}
all other primed variables equalling the unprimed ones. From
\THREETHREEFIVE\ we find that $-\infty<\lambda<\infty$ corresponds to
$0<\lambda'<\infty$. If we perform a path integral with the primed
variables and sum over both positive and negative $\lambda'$ then we are
summing over more than is being summed in the unprimed variable path
integral.

We thus see sources of ambiguity on the quantisation of the relativistic
particle working with a Fadeev-Popov approach in sec  3.2 and a BRST
approach in sec. 3.3. In fact the action we start with, \THREEFOUR , is
itself ambiguous because of the two possible signs of the square root. The
particle trajectory would keep switching in general between timelike and
spacelike, and at each switch we have to choose afresh  the sign of the
real or imaginary quantity obtained in these two cases respectively. This
suggests that the world line configuration should be described by the pair
$\{X^\mu(\tau),\sigma(\tau)\}$ with $\sigma=\pm 1$ giving the choice of
root. Evaluating the quadratic form of the action (given in \THREEEIGHT )
classically we find the sign of $\lambda$ to be related to the sign of the
square root chosen for \THREEFOUR .

The above discussion suggests a close connection between the ambiguities
found in three different approaches to the quantum relativistic particle,
and it would be good to determine if they indeed are the same. For the
rest of this paper we simply adopt as basic the picture of two complex
conjugate propagations on the world line, with switching between them
possible through the regulator matrix.

\subsec{Summary.}

We may describe our quantisation of the relativistic scalar
particle in the following colloquial terms. To obtain the
usual Feynman propagator the particle was allowed to
move both backwards and forwards in target space time.
But it moved only forwards in its proper time. Now we
are allowing the particle to move both backwards and forwards
in proper time also; this corresponds to flipping between the
two signs of the world line einbein.\foot{
I am indebted to A. Vilenkin for this way of
expressing the result.} Giving the amplitude to
reverse orientation of proper time amounts to specifying
an `exponential of quadratic' density matrix,. Putting
this amplitude to zero lets us recover the Feynman propagator,
by restricting to the positive sign of the einbein.

Thus we find a covariant description of the notion of
the notion of density matrix. We see that in the first quantised
dsecription, there is very little difference between causal
perturbation theory for pure states and for mixed states.

{\it Note:}\quad  $\tr \rho$ is finite only  for $4\alpha
\gamma <(1-e^{-\beta})^2$ (see appendix).
Consider the
initial value problem for  charged scalar field in a small box  with
a  (suddenly switched on) strong electric field. We may wish to expand the
field in modes which are eigenfunctions of the Hamiltonian.  The above
inequality
is violated at the point where the electric field becomes
strong enough to destabilise the vacuum \ref\FULL{S.A.
Fulling, `Aspects of Quantum Field Theory in Curved
Space-time' (1989) Cambridge University Press.}.  For stronger fields
the mode coefficients are $b$, $d^\dagger$, $[b,d^\dagger]=-i$
instead of $a$, $\ad$, $[a,\ad]=1$. We expect that black holes would also be
characterised by a similar violation.  The density matrix is still
`exponential of quadratic in the field' but \TWOSEVENPP\ is
not a useful representation. We will ignore this limitation  of the form
\TWOSEVENPP\ because for our examples it will be adequate.

\newsec{A curved space example: spacetime with expansion.}

Consider the free scalar field (\TWOONE\ with $\lambda=0$) propagating in
$1+1$ spacetime with metric
\eqn\FOURONE{ds^2~=~C(\eta)[d\eta^2-dx^2], \quad\quad -\infty<\eta<\infty,
\quad 0\le x<2\pi}
\eqn\FOURTWO{C(\eta)~=~A+B\tanh\kappa\eta, \quad\quad A>B\ge 0}
The conformal factor $C(\eta)$ tends to $A\pm B$ at $\eta\rightarrow  \pm
\infty$. The limit
$\kappa\rightarrow\infty$ gives a step function for $C(\eta)$; the
Universe jumps from scale factor $A-B$ to $A+B$ at $\eta=0$. We will work
in this limit to ensure simpler expressions.

For $\eta\rightarrow -\infty$ it is natural to expand $\phi$ as
\eqn\FOURTHREE{\phi(\eta,x)~=~\sum_{n-=\infty}^\infty {1\over
\sqrt{2\pi}}{1\over
\sqrt{2\omega_n^-}}(a_ne^{-i\omega_n^-\eta+inx}~+~a_n^\dagger
e^{i\omega_n^-\eta-inx} )} with
\eqn\FOURFOUR{\omega_n^-~=~(n^2+(A-B)m^2)^{1/2}>0}
We define the `in' vacuum by
\eqn\FOURFIVE{a_n|0>_{in}=0, \quad \quad {\rm for~~ all} ~~n}
Similarily, for $\eta\rightarrow \infty$ we write
\eqn\FOURSIX{\phi(\eta,x)~=~\sum_{n-=\infty}^\infty {1\over
\sqrt{2\pi}}{1\over
\sqrt{2\omega_n^+}}(\tilde
a_ne^{-i\omega_n^+\eta+inx}~+~\tilde a_n^\dagger e^{i\omega_n^+\eta-inx}
)} with
\eqn\FOURSEVEN{\omega_n^+~=~(n^2+(A+B)m^2)^{1/2}>0}
The `out' vacuum is defined through
\eqn\FOUREIGHT{\tilde a_n|0>_{out}=0, \quad \quad {\rm for~~ all} ~~n}
The `out' vacuum does not equal the `in' vacuum, even in the free theory:
\eqn\FOURTEN{|0>_{out}~=~C_0e^{{b_0\over 2}\ad_0\ad_0}\prod_{n>0} C_ne^
{b_n\ad_n\ad_{-n}}|0>_{in}}
\eqn\FOURELEVEN{b_n=-{\omega_n^+-\omega_n^-\over
\omega_n^++\omega_n^-},\quad\quad C_n=(1-b_n^2)^{1/2}, \quad\quad
n\ge 0.}

As mentioned in the introduction,  evaluating the analogue
of the Polyakov path integral gives
\RUMPF\ (we denote the pair
$(\eta, x)$ by $z$)
\eqn\FOURTWELVE{<z_2|\int_0^\infty d\lambda D[X]e^{-i\int_0^1
d\tau(1/2)(\dot X^2/\lambda+m^2\lambda)}|z_1>~=~{
{}_{out}<0|T[\phi(z_2)\phi(z_1)]|0>_{in}\over {}_{out}<0|0>_{in}}} (We
have integrated out $p$ in the phase space path integral.) It is not
surprising that both vacuua appear  in this quantity; after all the action
and measure are covariantly given and should not distinguish the past or
future as special. Using $-i$ in place of $i$ in the exponential gives
${}_{in}<0|\tilde T[\phi(z_2)\phi(z_1)]|0>_{out}/{}_{in} <0|0>_{out}$.
($\tilde T$ denotes anti-time-ordering.)

We wish to compute the propagator appropriate to the initial
value problem based on the `in' vacuum at $t=-\infty$.  This implies
a density matrix $\rho\equiv\rho_0=|0>_{in}{}_{in}<0|$.
To develop a perturbation theory using $\rho_0$ we need a real time
contour running from $\eta=-\infty$ to $\eta=\infty$, and then back to
$\eta=-\infty$ where we insert $\rho_0$ and take a trace to close the
path. This perturbation theory requires  a matrix propagator $D$:
\eqn\FOURTHIRTEEN{D(z_2,z_1)~=~\pmatrix{{}_{in}
<0|T[\phi(z_2)\phi(z_1)]|0>_{in} & {}_{in}<0|\phi(z_1)\phi(z_2)|0>_{in}\cr
{}_{in} <0|\phi(z_2)\phi(z_1)|0>_{in}& {}_{in}<0|\tilde
T[\phi(z_2)\phi(z_1)]|0>_{in}\cr}}

Our goal is to see if there exists a choice of regulator matrix
$M$ such that evaluting
\eqn\FOURIONE{D~=~ \int_0^\infty  \lambda
e^{-i\lambda H-\epsilon\lambda M}, \quad  H=\pmatrix
 {(\boxx+m^2)&0\cr 0&-(\boxx+m^2)\cr }}
gives the matrix propagator \FOURTHIRTEEN .

Let us discuss the general structure of the  world line
Hamiltonian and regulator matrix in \FOURIONE .
$H$ and $M$  have a `2x2 matrix  of operators'
 structure, on account of the
two signs of the einbein $\lambda$ possible on the world line.
The operator $\boxx+m^2$ acts on the space of all
functions of $(\eta,x)$. Formally,  each operator in
the 2x2 matrix $M$ also acts on this space, but as noted
at the end of section 3.1 the only space affected by
these operators is the space of `on shell' wavefunctions,
i.e., solutions of $(\boxx+m^2)f(\eta,x)=0$.  In the flat space
examples  \THREETWOI , \THREETHREE\ we had translational
invariance in space and time, so each operator could connect
one Fourier mode $p$ only to itself. Thus for each $p$,  $M$
was just a 2x2 matrix of  numbers. In the present example,
we retain  $x$-translational invariance, so  different $x$-Fourier modes
(labelled by $n$)  are not mixed by $M$. But time
translational invariance is broken by the expansion,  and so the two
solutions of the wave equation for fixed $n$ can be transformed
into each other by the elements of $M$.  Thus for each $n$ we need to consider
a 4x4 matrix, which acts on a
column vector $(\{f_{n}^1, f_{n}^2\}^+,\{f_{n}^1,f_{n}^2\}^-) $.
Here the first pair of functions propagate on the world line as $e^{i
(\boxx+m^2)}$
while the second pair propagates as $e^{-i(\boxx+m^2)}$.
Within each type ($+$ or
$-$) we have two linearly independent  solutions
of $(\boxx+m^2)f=0$.   Thus  for example the
element $M_{41}$ of the regulator matrix gives the
amplitude for the einbein to change sign from
positive to negative,  and for the  wavefunction on the world line
to change from $f_n^1$ to  $f_n^2$.

Let us set up the calculation of Green's functions in the first quantised
formalism. We need eigenfunctions of the world line Hamiltonian:
\eqn\FOURFOURTEEN{(\boxx+m^2)\psi_s(\eta,x)~=~-H\psi_s(\eta,x)~=~s
\psi_s(\eta,x)}
The following is a complete set: ($-\infty<n<\infty$) $${m^2+{n^2\over
A+B}<s<m^2+{n^2\over A-B}:\quad\quad\quad\quad\quad\quad\quad\quad\quad
}$$ $${\eqalign{\quad\quad\quad\psi_{\tilde \nu_+,n}(\eta ,x)~=
&~{e^{inx}\over \sqrt{2\pi}} e^{-\tilde \nu_+\eta},  \quad\eta>0 \cr
&{e^{inx}\over \sqrt{2\pi}}[{1\over 2}(1+{\tilde \nu_+\over i \nu_-})
e^{-i\nu_-\eta} + {1\over 2}(1-{\tilde \nu_+ \over i\nu_-})
e^{i\nu_-\eta}], \quad \eta <0 \cr}}$$
\eqn\FOURSIXTEEN{\tilde \nu_+=[-(A+B)(m^2-s)-n^2]^{1/2}~>~0, \quad
\nu_-=[(A-B)(m^2 -s)+n^2]^{1/2}~>~0}

$$ {-\infty<s<m^2+{n^2\over A+B}:\quad\quad\quad\quad\quad\quad\quad\quad
\quad}$$
$$ {\eqalign{\quad\quad\quad\psi_{\nu_+,n}(\eta ,x)~=&~{e^{inx}\over
\sqrt{2\pi}}
e^{- i\nu_+\eta},  \quad\eta>0 \cr &{e^{inx}\over \sqrt{2\pi}}[{1\over
2}(1+{ \nu_+\over  \nu_-}) e^{-i\nu_-\eta} + {1\over 2}(1-{ \nu_+ \over
\nu_-}) e^{i\nu_-\eta}], \quad \eta <0 \cr}}$$
\eqn\FOURSEVENTEEN{\nu_\pm^2=(A\pm B)(m^2-s)+n^2, \quad\quad
{\rm sign}(\nu_+)={\rm sign}(\nu_-)}

These functions are normalised as
\eqn\FOURSEVENTEENI{\eqalign{(\psi_{\tilde \nu_+',n'}, \psi_{\tilde
\nu_+,n} )~=~&\int d\eta dx C(\eta)\psi_{\tilde \nu_+',n'}^*(\eta,x)
 \psi_{\tilde
\nu_+,n}(\eta,x) \cr
=~&\delta_{n',n}\pi (A+B) {\nu_-^2+\tilde \nu_+^2
\over 2\nu_-\tilde \nu_+}
\delta (\tilde \nu_+'-\tilde \nu_+) \cr }}
\eqn\FOURSEVENTEENII{(\psi_{\nu_+',n'},
\psi_{\nu_+,n})~=~ \delta_{n',n}\pi (A+B) [{(\nu_-+\nu_+)^2 \over
2\nu_-\nu_+} \delta (\nu_+'-\nu_+) ~+~ {\nu_-^2-\nu_+^2 \over 2
\nu_-\nu_+} \delta (\nu_+'+\nu_+)]}

Let us first recover the propagator \FOURTWELVE\ in this formalism. Let
$\eta', \eta>0$. The range $m^2+{n^2\over A+B}<s<m^2+{n^2\over A-B}$ gives
the contribution
\eqn\FOURTWENTYTHREE{\eqalign{\sum_n&\int_{\tilde \nu_+,\tilde \nu_+'=0}
^{\sqrt{2B/(A-B)}|n| } d\tilde \eta_+' d\tilde \eta_+
<\eta',x'|\psi_{\tilde\nu_+,n}> <\psi_{\tilde\nu_+,n}| \int_0^\infty
d\lambda e^{i(-s+i\epsilon)\lambda} |\psi_{\tilde
\nu_+',n}>\cr
&\quad\quad\quad\quad\quad\quad\quad\quad\quad\quad\quad\quad
<\psi_{\tilde\nu_+',n}|\eta,x> \delta(\tilde\nu_+'-\tilde\nu_+)[(
A+B)\pi]^{-1}{2\nu_-\tilde \nu_+ \over \nu_-^2+\tilde \nu_+^2}\cr
&=\sum_n{e^{in(x'-x)}\over 2\pi}
\int_0^{\sqrt{2B/(A-B)}|n|} d\tilde \nu_+ ({-i\over \pi})
e^{-\tilde\nu_+(\eta+\eta')} {2\nu_-\tilde\nu_+\over
\nu_-^2+\tilde\nu_+^2 } {1\over (\tilde\nu_+^2+n^2+m^2(A+B))}\cr}}

Similarily the range $-\infty<s<m^2+{n^2\over A+B}$ provides the
contribution
\eqn\FOURTWENTYFIVE{\sum_n{e^{in(x'-x)}\over 2\pi}
\int_{-\infty}^\infty d \nu_+
{i\over 2\pi} [e^{i\nu_+(\eta'-\eta)}~+~{\nu_+-\nu_-\over \nu_++\nu_-}
e^{i\nu_+(\eta'+\eta)}] {1\over (\nu_+^2-n^2-m^2(A+B)+i\epsilon)}} There
is a branch cut in the complex $\nu_+$ plane joining $\nu_+=\pm
i\sqrt{2B\over A-B}|n|$. The $\nu_+$ integral in \FOURTWENTYFIVE\ has a
discontinuous jump across this cut for the part multiplying
$e^{i\nu_+(\eta'+\eta)}$. The contribution from \FOURTWENTYTHREE\ can be
added to \FOURTWENTYFIVE , however, with the identification  $\tilde
\nu_+=i\nu_+$. This results in a contour passing over the cut. Evaluating
the resulting contour integrals one obtains the result
\eqn\FOURTWENTYSIX{D^0_{11}
(\eta',x',\eta,x)~=~\sum_n{e^{in(x'-x)}\over 2\pi} {1\over
2\omega^+_n}e^{-i\omega_n^+|\eta'-\eta|}~+~{1\over 2\omega_n^+}
{\omega_n^+- \omega_n^- \over \omega_n^++\omega_n^-}
e^{-i\omega_n^+(\eta'+\eta)}, ~ {\rm for} ~~\eta, \eta'>0} which may be
readily verified in the operator language using \FOURSIX\ and
\FOURTEN .

 We now proceed to the `in-in' matrix propagator. From the expansion of
 the field operator $\phi$ in creation and annihilation modes, we can
 readily compute each element of the matrix propagator.  We try
 to find a regulator matrix $M$ which will reproduce this propagator
 by the first quantised calculation. To display the
 4x4 matrix $M$, we need to choose a basis in the space of solutions
 of the wave equation for each $x$-Fourier mode $n$. We choose the basis
($\omega_n^->0$)
\eqn\FOURTWENTYEIGHT{\eqalign{f^1_n~=~&{e^{inx}\over
\sqrt{2\pi}}e^{-i\omega_n^- \eta}, \quad \quad \eta<0\cr
=~&{e^{inx}\over \sqrt{2\pi}}[{1\over 2}(1+{\omega_n^-\over \omega_n^+})
e^{-i\omega_n^+\eta} ~+~ {1\over 2}(1-{\omega_n^-\over
\omega_n^+})e^{i\omega_n^+ \eta}],\quad\eta>0 \cr
f_n^2(\eta,x)~=~&f_n^{1*}(\eta,x) \cr }}

The result we get is that the required propagator is obtained for
\eqn\FOURTWENTYNINE{M~=~\pmatrix{1&0&0&0\cr 0&1&{4B_n\over 1+4B_n^2}&
{-2\over 1+4B_n^2}\cr {-2\over 1+4B_n^2}&{4B_n\over 1+4B_n^2}&
1&0&\cr0&0&0&1\cr}} where
\eqn\FOURTHIRTYONE{B_n~=~-{1\over 2}{\omega_n^+-\omega_n^-\over
\omega_n^++\omega_n^-} }
(In this computation we need to note that there are
many equivalent descriptions of the delta function. For example ${\rm
lim}_{\epsilon\rightarrow 0}{\epsilon\over x^2+\epsilon^2}=\pi\delta(x)$,
but ${\rm lim}_{\epsilon\rightarrow 0}{\epsilon\over (x+i\epsilon)^2}=0$;
thus any amount of the latter can be added to the former to get
a different  representation of the $\delta$-function.)

 As was the case in the examples of the last section, $M$ does not
 have enough independent entries to reproduce any arbitrary
  2x2 matrix of propagators. In the present example any propagator has the
freedom of 4 constants for each $x$-Fourier mode $n$: we can add
$\sum_{i,j=1}^2 C_{ij}f^{(i)}(z)f^{(j)}(z')$ where $f^{(i)}$ are given in
\FOURTWENTYEIGHT .  A 2x2 matrix  of propagators thus has
16 free parameters for each $n$.  But the entries  in the diagonal
2x2 blocks of $M$ are not  all independent, in the sense that  some changes to
$M$ do not affect the final propagator obtained. (For example,  scaling
$M\rightarrow cM$ leaves the propagator unchanged.) Thus  at first it seems a
coincidence that we could find
$M$ to obtain \FOURTHIRTEEN .

To understand this coincidence we recall that \FOURTHIRTEEN\ came from a
`closed time path' formulation. This formulation is really covariant
(and not dependent on any choice of `time'). Each spacetime point is
covered twice in the description (alternatively we can say that there is a
doubling of fields \ref\UMEZ{H. Umezawa, H. Matsumaoto and M Tachiki,
`Thermo Field Dynamics and Condensed States' (1982) North Holland.}).
Over this double cover of spacetime we are looking for  a first quantised
description of the propagator. Clearly such a propagator can be represented by
the usual path integral in each cover, together with an amplitude to switch
between covers. Thus all matrix propagators
arising in the closed time path formalism will have the special form
that arises from allowing orientation changes on the world line in the first
quantised language.

\newsec{Strings.}

Let us now come to strings. Requiring consistent propagation of the
string determines the field equations for the background on
which the string moves. But use of `in-out' and `in-in' propagators
might give diffrerent field equations, as we argue in this section.

To obtain the necessary ingredients for the argument, we first
consider a simple case: strings in a contant temperture bath, in
a time independent background.  Since
we are at constant temperature, we can use for the moment
the periodic imaginary time trick to
take into account the temperature  \ref\SAT{B. Sathiapalan, Phys. Rev.  Lett.
{\bf 58} (1987) 1597, J.J. Attick and E. Witten, Nucl. Phys. {\bf B 310}
(1988) 291.}.

We expect that the
vanishing  $\beta$-function conditions will reproduce, in some
approximation, the Einstein equation
\eqn\FIVEONE{G_{\mu\nu}~=~<T_{\mu\nu}>}
where $T_{\mu\nu}$ is the stress-tensor of the thermal distribution.
The tree level  $\beta$-function calculation is a local  one: local on the
world
sheet and local in target space.  At tree level we thus obtain
$G_{\mu\nu}=0$;  we are unable to `see' the thermal bath.  The
$\beta$-function also has a  contribution at one loop in
the string coupling expansion: we must consider tadpoles attached
to the world sheet by necks thinner than the world sheet cut-off scale
\ref\FS{W. Fishler and L. Susskind, Phys. Lett. {\bf B171} (1986)
383, {\bf B173} (1986) 262.} . The loop of the tadpole can wind around the
compact
imaginary time direction. This  takes into account the value of
  temperature and produces the contribution  $<T_{\mu\nu}>$
  required for the field equation \FIVEONE\  \ref\HELL{M. Hellmund
and J. Kripfganz, Phys. Lett. {\bf B241} (1990) 211.}.

  Let us now ask how the  $\beta$-function equations would reproduce
  \FIVEONE\  if we are {\it not} permitted use of the periodic
  imaginary time trick to take into account the particle distribution.
  The question is pertinent, because as mentioned in the introduction
  time independent distributions are unnatural in a theory of gravity,
  and any time dependence would invalidate the periodic imaginary
  time trick. There is  no
  `analytic continuation of time' for a general geometry.
  \foot{It is possible to study some time dependent
  situations by making a canonical transformation on the target
  space co-ordinates $X^\mu$ and taking the new $X^0$ to be
  imaginary and compact; this does not however have the
  physical appeal of the real-time method.}

  In the absence of the imaginary time winding mode, neither
  the tree nor the loop  contributions to the $\beta$-function  see the
  thermal distribution.  But considering the extended propagator-vertex
  structure of the real time formalism we should be able to recover
  the RHS of \FIVEONE . The one loop tadpole is composed
  of two propagators and two vertices (including the vertex
  joining the tadpole to the rest of the world sheet).  These vertices
  can be type 1 or type 2,  in the notation of section 2, and the
  propagators will be the corresponding elements of the 2x2 matrix
  of propagators.  These propagators depend on the ensemble
  distribution function, and so `see' the temperature.  It would be
  interesting to explicitly carry out this calculation, but here we
  will just assume that the real time method will work for strings
  the same way it does with particles.  Note that such a calculation does
  not need the restriction to time-independent distribution functions.

 We now argue, again with
 the scalar field analogy,  that there is a direct connection between the
 choice of propagator and the value of $<T_{\mu\nu}>$ seen by
 the $\beta$-function calculation for the string background.
 Consider the scalar field in flat space and ignore the
 mass term for the moment (since the string case has massless
 fields on the world sheet). Then Einstein's equations read
 \eqn\FIVEIPONE{R_{\mu\nu}(x)~=~<\partial_\mu\phi(x)
 \partial_\nu\phi(x)>~=~\lim_{x\rightarrow x'}\partial_\mu
 \partial'_\nu <\phi(x)
 \phi(x')>}
(We ignore any subtraction in the definition of $T_{\mu\nu}$
since the string case is ultraviolet finite and has no subtraction.)
Using the analogue of the Polyakov path integral
\eqn\FIVEIPTWO{<\phi(x)
 \phi(x')>=\int_0^\infty d\lambda D[X]_{X(0)=x}^{X(1)=x'}e^{-i\int_0^1
d\tau(1/2) X,_{\tau}^2/\lambda}= \int_0^\infty d\lambda
D[X]_{X(0)=x}^{X(\lambda)=x'}e^{-i\int_0^1
ds(1/2) X,_{s}^2}}
where $ds=\lambda d\tau$ is the length  element along the world line.
Then
\eqn\FIVEIPTHREE{\partial_\mu\partial'_\nu<\phi(x)
 \phi(x')>~=~ \int_0^\infty d\lambda D[X]_{X(0)=x}^{X(\lambda)=x'}{dX_\mu\over
ds}(0){dX_\nu\over
 ds}(\lambda)e^{-i\int_0^\lambda
ds(1/2) X,_{s}^2}}
Let us compare the above with the $\beta$-function equation for
string theory in  a finite
temperature bath described in
 the imaginary-time formalism \HELL :
 \eqn\FIVEIPFOUR{ \beta_{\mu\nu}~=~R_{\mu\nu}~+~2\nabla_\mu
 \nabla_\nu~=~{e^{2\phi}\over 2\pi\beta V d}\sum_{g\ge 1}
 <\partial X_\mu\bar\partial X_\nu>_g}
 ($\phi$ is the dilaton, $V$ is the voloume of spacetime, $d$
 is the number of space dimensions.)
 The RHS of \FIVEIPFOUR\ is the analogue of \FIVEIPTHREE .
(The factor $(\beta V)^{-1}$ in \FIVEIPFOUR\ compensates
 for the $X^\mu$ zero mode; without this compensation we would compute the
stress tensor not  at a point $x$ but instead integrated over all the
 range of coordinates $X^\mu$.)  Examining the derivation of
 \FIVEIPTHREE , we expect that in the string case   also the stress tensor
contribution
 to the  $\beta$-function comes in a direct way  from the propagator of the
first quantised
 language.  (Thus in \FIVEIPFOUR\ it comes from the propagator
 in spacetime with a compact Euclidean time direction.) If we
 have a general time dependent spacetime (with Minkowski
 signature) and we compute amplitudes by the Polyakov
 prescription, then the propagator is \hfill\break
  ${}_{out}<0|T\{\phi(x)\phi(x')\}|0>_{in}/{}_{out}<0|0>_{in}$ and so we
 we expect to incorporate \hfill\break
  ${}_{out}<0| T_{\mu\nu}|0>_{in}/{}_{out}<0|0>_{in}$
 in the background field equations obtained
 {}from the  $\beta$-function  constraint.
 This is the wrong kind of object to have
 in the  gravity equation, as we illustrate with the following
 simple example.

  We consider the free scalar field on the  $1+1$ spacetime
 \FOURONE , \FOURTWO . We compute  `in-out' vacuum expectation values and
`in-in' vacuum  expectation values,  for the stress tensor.
 To achieve a parallel with the string case, we consider no
 subtraction in defining $T_{\mu\nu}$, but compute the
 contribution from different field modes separately, to get
 finite answers. (We can imagine an ultraviolet cutoff regulates
 the sum over different modes, to make sense of  sum over modes.)
   More specifically, we compute the contribution to
\eqn\FIVETWO{\eqalign{<T_{\mu\nu}(\eta)>_{in~in}~\equiv~&
\int_{x=0}^{2\pi} dx~ {}_{in}<0|T_{\mu\nu}(\eta,x)|0>_{in} \cr
 <T_{\mu\nu}(\eta)>_{in~out}~\equiv~&
\int_{x=0}^{2\pi} dx ~{}_{out}<0|T_{\mu\nu}(\eta,x)|0>_{in}
/{}_{out}<0|0>_{in} \cr }}
{}from the field modes with $x$-Fourier component $\pm n$:
 \eqn\FIVETHREE{
\phi_{\pm n}~=~a_{\pm n}~f_{\pm n}~+~
 \ad_{\pm n}~f^*_{\pm n}}
\eqn\FIVEFOUR{\eqalign{f_{\pm n}~=~&{e^{\pm i n x}\over \sqrt{
2 \pi}\sqrt{2 \omega_-}}e^{-i\omega_-\eta}\quad\quad , \eta<0 \cr
=~&{e^{\pm i n x}\over \sqrt{
2 \pi}\sqrt{2 \omega_-}}[{\omega_++\omega_-
\over 2 \omega_+}e^{-i\omega_+\eta}~+~
{\omega_+-\omega_-
\over 2 \omega_+}e^{i\omega_+\eta}]\quad , \eta>0 \cr}}

The stress tensor for the scalar field is
\eqn\FIVEFIVE{T_{\mu\nu}~=~{1\over 2}(\partial_\mu\phi
\partial_\nu\phi+\partial_\nu\phi \partial_\mu\phi) -{1\over 2}
g_{\mu\nu}\partial_\lambda\phi\partial^\lambda\phi+{1\over 2}
m^2\phi^2}
Define
\eqn\FIVESIX{T_{\mu\nu}[f,g]~=~{1\over 2}(\partial_\mu f
\partial_\nu g+\partial_\nu f \partial_\mu g) -{1\over 2}
g_{\mu\nu}\partial_\lambda f\partial^\lambda g+{1\over 2}
m^2 f g}

The expansion \FIVETHREE\ is in terms of the `in' vacuum
creation and annihilation operators. We recall that
\eqn\FIVESIXP{|0>_{out}~=~Ce^{b\ad_n\ad_{-n}}|0>_{in}}
where $b=-(\omega_+-\omega_-)/(\omega_++\omega_-)$.
We have
\eqn\FIVESEVEN{\eqalign{<T_{00}>_{in~in}~=&~T_{00}[
f_n,f^*_n]+T_{00}[
f_{-n},f^*_{-n}] \cr
<T_{11}>_{in~in}~=&~T_{11}[
f_n,f^*_n]+T_{11}[
f_{-n},f^*_{-n}] \cr
<T_{00}>_{in~out}~=&~T_{00}[
f_n,f^*_n]+T_{00}[
f_{-n},f^*_{-n}]+b\{ T_{00}[
f^*_n,f^*_n]+T_{00}[
f^*_{-n},f^*_{-n}]\}\cr
<T_{11}>_{in~out}~=&~T_{11}[
f_n,f^*_n]+T_{11}[
f_{-n},f^*_{-n}]+b\{ T_{11}[
f^*_n,f^*_n]+T_{11}[
f^*_{-n},f^*_{-n}]\}\cr}}
($T_{01}$, $T_{10}$ vanish.)

{}From \FIVEFOUR\ we get
\eqn\FIVEEIGHT{\eqalign{<T_{00}>_{in~in}~=~&\omega_- ,\quad
\eta<0 \cr
=~&\omega_++2\omega_+<N_{out}>, \quad \eta>0 \cr
<T_{11}>_{in~in}~=~&n^2/\omega_- ,\quad
\eta<0 \cr
=~&{n^2\over 2 \omega_-\omega_+^2}(\omega_+^2+
\omega_-^2)-{m^2\over 2\omega_+^2}(\omega_+^2-\omega_-^2)
{\rm cos}(2\omega_+ t), \quad \eta>0 \cr
<T_{00}>_{in~out}~=~&\omega_- ,\quad
\eta<0 \cr
=~&\omega_+, \quad \eta>0 \cr
<T_{11}>_{in~out}~=~
&{n^2\over \omega_-} - {(\omega_-^2-n^2)\over \omega_-}({\omega_+-\omega_-
\over \omega_++\omega_-})e^{2i\omega_- \eta} ,\quad
\eta<0 \cr
=~&{n^2\over \omega_+} - {(\omega_+^2-n^2)\over \omega_+}({\omega_+-\omega_-
\over \omega_++\omega_-})e^{-2i\omega_+ \eta}, \quad \eta>0 \cr}}
Here $<N_{out}>=(\omega_+-\omega_-)^2/4\omega_+\omega_-$
is the number of `out' particles in the `in' vacuum.

We see that  $<T_{00}>_{in~in}$ has a direct physical
interpretation. For $\eta<0$ (before expansion) we are in the `in' vacuum
state, and we just get the vacuum energy $\omega_-/2$ for
each of the two modes $n,-n$ considered here. For
$\eta>0$ (after expansion) we get the vacuum energy
$\omega_+/2$ for each mode, plus the energy of created
`out' particles.

$<T_{00}>_{in~out}$ on the other hand gives only the vacuum
energies; there is no contribution from created particles.
Worse,  $<T_{11}>_{in~out}$ is complex, and so
cannot be on the RHS of Einstein's equation for the gravitational
field.

This result is not surprising; we know that in the `in-out' formalism
the effective field equation    for
a real field can become
complex \PAZ . The `in-in' formalism, by contrast, is causal
and yields real expectation values for self-adjoint operators.
Thus it appears essential to extend the first quantised
path integral prescription for strings  so that we can get the 2x2 matrix
propagators of the real time formalism. Extending the discussion of
section 3 to the string case we would obtain a sum over
both signs of the zwiebein on the world sheet (instead of
just a sum over metrics.) The amplitude to flip sign of the
zwiebein would encode a density matrix which describes
`exponential of quadratic' string distributions (initially
present or `created'.) Because the string is two dimensional, there is an
interesting variety of ways that the zweibein can change sign,
but for the purposes of the present paper we can just assume
that the string is expanded into particle modes, in which case
the above approach for particles applies.

{\it Note:}\quad Leblanc \ref\LEB{Y. Leblanc, Phys. Rev. {\bf D 36} (1987)
1780, {\bf D 37}
(1988) 1547, {\bf D 39} (1989) 1139.} studied the real time formalism for
open and closed strings, for the case of constant temperature in flat space.
The
propagator was computed in the `thermo-field dynamics' language, which used
the Niemi-Semenoff time path, and so was given by \THREETWO ,
\THREETWOI . For this time-independent situation amplitudes were
computed and the Hagedorn temperature recovered. To extend the
string calculations to  arbitrary spacetime
geometry, as required in our approach, needs further
progress in defining and computing amplitudes in Minkowski signature
spacetimes. We will discuss this issue in more detail elsewhere.

\newsec{Discussion.}

Let us summarise the arguments and results of this paper. Requiring
consistent propagation for the first quantised string  should
determine the background geometry. Do the background field
equations incorporate the stress tensor of `created strings' in time
dependent geometries? To investigate this question we studied
a simpler theory in first quantised language: the relativistic scalar
particle. From the example of strings at constant temperature, we
expect that backreaction from particle fluxes will show up
at one string loop  level in the  $\beta$-function calculation. But
this one loop contribution computes the stress tensor as two
derivatives of the propagator, and so the backreaction
seen by the  $\beta$-function calculation depends on the
choice
of propagator.  The Polyakov prescription of
summing over target space co-ordinates $X^\mu$ and world sheet
metrics gives an `in-out' vacuum propagator. Using the scalar field analogy,
we then argue that the gravity field equations obtained at
one loop level will have
an object like ${}_{out}<0| T_{\mu\nu}|0>_{in} /{}_{out}<0|0>_{in}$
for the backreaction. We demonstrated by a simple example
that such quantities are in general not real, and not the correct ones
to have in the field equation. We need instead a `true expectation
value' of the stress tensor, which will be achieved if we
have a   causal `in-in' propagator, instead of the `in-out' propagator.

In achieving a causal formulation of perturbation theory
we need to extend the propagator to a 2x2 matrix of propagators \PAZ .
This closed time path formulation (called the real time formulation
in non-equilibrium many body theory) also naturally handles
perturbation theory in the presence
of particle fluxes. In fact the perturbation scheme is built around
precisely the same kind of fluxes (exponential of quadratic) which
are obtained in particle creation effects.

Accepting the necessity of a  real time formalism for strings, we are
faced  with the question: can this formalism be obtained
in a natural way in a first quantised language?  We start
with the square root action of the scalar particle, and try to
reach the quadratic form that is the analogue of the Polyakov action.
We find that the proper time in the latter description does not
necessarily go forward. We need to supply the amplitudes
for any state  $|\psi>$ on the world line to connect to any
other state $|\psi'>$, when the proper time reverses
direction.  Setting all these amplitudes
to zero is a special choice that  forbids reversal
of proper time; choosing the forward direction of
this time reproduces the `in-out'
vacuum propagator.

But it is also possible to set these amplitudes to be nonzero.
With a particular choice of the amplitudes we found the matrix
propagator for perturbation  theory in a thermal bath. With
another choice we found the causal `in-in' propagator for a Universe
which is  in the
`in' vacuum but where expansion creates a bath of `out' particles.
In both these cases the real time description involved a density
matrix of the kind \TWOSEVENPP . For the thermal bath we have
$\alpha=\gamma=0$,
$\beta$ finite. For the `created particles' case we had a limit
for the coefficients in which a pure state density matrix was obtained.

For strings, the passage from the Nambu-Goto action to  a
quadratic action would bring in two signs  of world sheet  zweibien;
restricting to one sign would give the Polyakov amplitude.  As remarked at the
end
of section 3.1, only infinitely long world lines contribute
to the orientation flip amplitude of the world line. The analogue
of this for the string case would be that the effect of $\rho$
is felt only through the boundary of the moduli space of Riemann surfaces,
where a homologically trivial or non-trivial cycle is pinched.
A $\beta$-function calculation for the
string world sheet theory would have to take into account such pinches
while considering the small handle contribution studied by Fishler and
Susskind.
 Such a calculation would yield  a relation
between the classical fields and the particle fluxes, rather than just
among the classical fields giving the background. (The constant temperature
case quoted in section 5 is an example of this.)
The classical field strength describes a coherent state; the flux
describes an `exponential of quadratic' distribution of particles. The
natural way in which field and fluxes have been mixed in our
formalism leads us to expect that   in
a theory of gravity `effective equations' of the quantum theory
should have both these components, instead of having field values alone.

It would be interesting to obtain a non-perturbative treatment
(like a matrix model approach) for a situation like an evaporating black hole
where the backreaction from a flux is important in determining the evolution
of the geometry. Based on the discussion of this paper, it appears that
for this problem we should be looking not for  a conformal theory
but an `extended  conformal theory' (with both orientations of the world sheet
allowed). The conformal invariance condition would set up
a relation between the coherent field (contributing to the
$\beta$-function from tree level onwards) and the  orientation reversing
amplitude representing `string flux' (which contributes to the
$\beta$-function from one string loop onwards). The conformal theory
describes just one string, and so should be an inadequate description
of this situation.

\bigskip
\bigskip
\centerline{\bf Acknowledgements}
\bigskip
I would like to thank for helpful discussions R. Brooks, M. Crescimanno,
J. Cohn, S.R. Das, E. Farhi,  L. Ford, D. Freedman,  S. Gutman, J. Halliwell,
R. Jackiw, D. Jatkar, S.
Jain, K. Johnson, E. Keski-Vakkuri,  G. Lifshitz, S. Mukhi, M. Ortiz, A. Sen,
C. Vafa, A. Vilenkin and B. Zwiebach.
This work is supported in part by DOE grant DE-AC02-76ER.

\vfill
\eject

\appendix{A}
{Wick theorem for `exponential of quadratic' density matrices.}

We wish to establish Wick's theorem for density matrices of the form
\eqn\AONE {\rho ~=~e^{\gamma a a }e^{\alpha a^\dagger
 a^\dagger}e^{-\beta a^\dagger a}}

A string of creation and annihilation operators can be brought to normal
ordered form in the same way as for the usual Wick theorem in the vacuum.
What we need to show in addition is that
\eqn\ATWO{{1\over \Tr \rho}\Tr\{\rho \ad \dots \ad a \dots a\}~=~\sum
{\Tr\{\rho\ad\ad\}\over \Tr\rho}\dots {\Tr\{\rho\ad a \}\over
\Tr \rho}\dots {\Tr \{\rho a a\} \over \Tr \rho}} where the RHS has a
summation over all possible pairings of the $\ad$, $a$ operators on the
LHS. We sketch below some of the steps involved in the derivation.

Note
\eqn\ATHREE{e^{\alpha \ad \ad}e^{-\beta \ad a}~=~
e^{-\beta  \ad a}e^{\alpha'\ad \ad}} with $\alpha'=\alpha e^{2\beta}$. A
straightforward calculation gives
\eqn\AFOUR{\Tr \rho~=~(1-e^{-\beta})^{-1}[1-{4\alpha\gamma\over
(1-e^{-\beta})^2} ]^{-1/2}} which we may rewrite as
\eqn\AFIVE{
\Tr \rho ~=~ \Tr \{e^{\gamma a a }e^{-\beta \ad a }e^{\alpha'\ad\ad}\}
{}~=~(1-e^{-\beta})^{-1}[1-{4\alpha'\gamma\over (1-e^{-\beta})^2} ]^{-1/2}}
{}From \AONE\ we see that in \ATWO\ there must be either an even number
$(2p)$ of `$a$' oscillators and an even number $(2q)$ of `$\ad$'
oscillators, or an odd number $(2p+1)$ of `$a$' and an odd number $(2q+1)$
of `$\ad$' oscillators. Assume first that we have the former case. Then
the LHS of
\ATWO\ is obtained as
\eqn\ASIX{\Tr \{ \rho (\ad)^{(2q)}(a)^{2p}\}~=~{1\over \Tr \rho}
(\partial_{\alpha'})^q(\partial_\gamma)^p \Tr \rho} (Partial derivatives
are taken with $\alpha'$, $\beta$, $\gamma$ as independent variables,
unless otherwise mentioned.) In particular,
\eqn\ASEVEN{\eqalign{
<\ad\ad>~=&~{1\over \Tr \rho}\partial_{\alpha'}\Tr \rho ~=~2\gamma
e^{-2\beta}/K~
\equiv~A \cr
<aa>~=&~{1\over \Tr \rho}
\partial_\gamma \Tr \rho ~=~2\alpha'e^{-2\beta} /K ~\equiv~B \cr
<\ad a >~=&~-{1\over \Tr \rho} \partial_\beta [\Tr
\rho]_{\alpha,\gamma}~=~ e^{-\beta}(1-e^{-\beta})/K ~\equiv~C \cr}} where
in computing $C$, $\partial_\beta$ is a partial derivative with $\alpha$,
$\gamma$ held fixed. Here
\eqn\AEIGHT{K~=~(1-e^{-\beta})^2-4\alpha'\gamma e^{-2\beta} }
We find
\eqn\ANINE{\Tr \rho~=~e^\beta[C^2-AB]^{1/2}~=~K^{-1/2}}
For fixed $\beta$
\eqn\ATEN{\eqalign{\partial_{\alpha'} A~=~2A^2 &\quad\quad
 \partial_\gamma A~=~2C^2 \cr
\partial_{\alpha'} B~=~2C^2 & \quad\quad\partial_\gamma B~=~2B^2 \cr
\partial_{\alpha'} C~=~2AC & \quad\quad
\partial_\gamma C~=~2BC \cr
}}

Using the above formulae, we can establish \ATWO\ by induction. Suppose
\ATWO\ holds with $2p$ operators `$a$' and $2q$ operators `$\ad$'. A
typical term on the RHS would have the form $FA^{n_1}B^{n_2}C^{n_3}$,
where $F$ is a constant and $n_1$, $n_2$, $n_3$  $\geq 0$. To establish
the result for $2p$ operators `$a$' and $2q+2$ operators `$\ad$' we get
for the LHS of \ATWO:
\eqn\AELEVEN{\eqalign{&{1\over e^\beta(C^2-AB)^{1/2}}\partial_\alpha'
[e^\beta
(C^2-AB)^{1/2}FA^{n_1}B^{n_2}C^{n_3}]~=~F[A^{n_1+1}B^{n_2}C^{n_3}\cr & ~+~
2n_1A^{n_1+1}B^{n_2}C^{n_3}
{}~+~2n_2A^{n_1}B^{n_2-1}C^{n_3}~+~2n_3A^{n_1+1}B^{n_2}C^{n_3}]}} The first
term on the RHS of \AELEVEN\ gives the pairing of the two new operators
$\ad$ with each other. The second term gives the $n_1$ ways to choose an
existing pair $(\ad\ad)$ and to contract the new $\ad$ operators with
members of this pair instead. The third term corresponds to choosing an
$(aa)$ pair in the original expression and contracting the `$a$' operators
with the new `$\ad$' operators instead. The last term corresponds to
exchanging the $\ad$ in an existing $\ad a $ pair with one of the new
$\ad$ operators.  It is easily seen that this generates all the new terms
required on the RHS of \ATWO\ for the induction to hold.

To work with the case of an odd number of $a$ and $\ad$ operators we start
with the expression $C\Tr \rho  = \Tr\{
 e^{\gamma a a }e^{-\beta\ad a } e^{\alpha'\ad\ad} \ad a\} $ in place of
$\Tr \rho$, and proceed as above to introduce extra $\ad\ad$  and $aa$
pairs in the induction.

\bigskip
\listrefs
\bye